\documentclass{article}

\usepackage{arxiv}

\usepackage[utf8]{inputenc} 
\usepackage[T1]{fontenc}    
\usepackage{hyperref}       
\usepackage{url}            
\usepackage{booktabs}       
\usepackage{amsfonts}       
\usepackage{amsmath}
\usepackage{nicefrac}       
\usepackage{microtype}      
\usepackage{cleveref}       
\usepackage{subcaption}    
\usepackage{lipsum}         
\usepackage{graphicx}
\usepackage{natbib}
\usepackage{doi}

\usepackage{booktabs}
\usepackage{makecell}
\usepackage{amsmath}

\usepackage[dvipsnames]{xcolor}\usepackage[draft,inline,nomargin]{fixme} \fxsetup{theme=color} 
\FXRegisterAuthor{cg}{cgg}{\color{RoyalBlue}CGG}
\FXRegisterAuthor{st}{ast}{\color{Maroon}AST}

\title{Linking the “inner” and “outer” self to mental health and brain networks}

\newif\ifuniqueAffiliation

\ifuniqueAffiliation 
\author{ \href{https://orcid.org/0000-0000-0000-0000}{\includegraphics[scale=0.06]{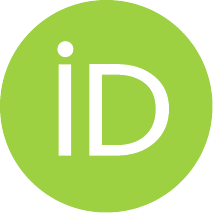}\hspace{1mm}David S.~Hippocampus}\thanks{Use footnote for providing further
		information about author (webpage, alternative
		address)---\emph{not} for acknowledging funding agencies.} \\
	Department of Computer Science\\
	Cranberry-Lemon University\\
	Pittsburgh, PA 15213 \\
	\texttt{hippo@cs.cranberry-lemon.edu} \\
	\And
	\href{https://orcid.org/0000-0000-0000-0000}{\includegraphics[scale=0.06]{orcid.pdf}\hspace{1mm}Elias D.~Striatum} \\
	Department of Electrical Engineering\\
	Mount-Sheikh University\\
	Santa Narimana, Levand \\
	\texttt{stariate@ee.mount-sheikh.edu} \\
}
\else
\usepackage{authblk}

\newbox{\orcid}\sbox{\orcid}{\includegraphics[scale=0.06]{orcid.pdf}} 
\author[1, 2]{%
	\href{https://orcid.org/0009-0006-5649-0872}{\usebox{\orcid}\hspace{1mm}Cosimo Agostinelli}%
}
\author[3]{%
	\href{https://orcid.org/0009-0003-6168-7459}{\usebox{\orcid}\hspace{1mm}Ivan Casanovas}%
}
\author[4]{%
	\href{https://orcid.org/0009-0000-7556-8037}{\usebox{\orcid}\hspace{1mm}Lochan Chaudhari}%
}
\author[5]{%
	\href{https://orcid.org/0009-0008-1006-4688}{\usebox{\orcid}\hspace{1mm}Arda Ergin}%
}
\author[6,7]{%
	\href{https://orcid.org/0009-0004-5697-5672}{\usebox{\orcid}\hspace{1mm}Pablo Estévez-Gutiérrez}%
}
\author[1]{%
	\href{https://orcid.org/0009-0006-3602-1593}{\usebox{\orcid}\hspace{1mm}Akanksha Gupta}%
}
\author[8]{%
	\href{https://orcid.org/0000-0002-9199-8237}{\usebox{\orcid }\hspace{1mm}Juliane T. Moraes}%
}
\author[9]{%
	\href{https://orcid.org/0009-0006-6509-4425}{\usebox{\orcid}\hspace{1mm}Mario Edoardo Pandolfo}%
}
\author[10]{%
	\href{https://orcid.org/0000-0003-0193-3067}{\usebox{\orcid}\hspace{1mm}Carlos Gershenson}%
}
\author[11]{%
	\href{https://orcid.org/0000-0002-7422-3421}{\usebox{\orcid}\hspace{1mm}Haily Merritt}%
}
\author[11]{%
	\href{https://orcid.org/0000-0002-2758-1891}{\usebox{\orcid}\hspace{1mm}Andreia Sofia Teixeira}%
}

\affil[1]{Aix-Marseille University, Marseille, France}
\affil[2]{Université de Toulon, CNRS, CPT,
Turing Center for Living Systems, 13009 Marseille, France}
\affil[3]{Departament de la Matèria Condensada, Universitat de Barcelona, Martí i Franquès, 1, Barcelona, Spain}
\affil[4]{Network Science Institute, Northeastern University Boston, USA}
\affil[5]{Universiteit van Amsterdam, Amsterdam, the Netherlands}
\affil[6]{ Institut d’Investigacions Biomèdiques August Pi i Sunyer (IDIBAPS), Barcelona, Spain}
\affil[7]{Universitat de Barcelona, Barcelona, Spain}
\affil[8]{Universitat Politècnica de Catalunya, Departament de Fisica, Terrassa, Spain}
\affil[9]{DIAG Department, Sapienza University of Rome, via Ariosto 25, Rome, Italy}
\affil[10]{School of Systems Science and Industrial Engineering, Binghamton University, USA}
\affil[11]{BRAN Lab, Network Science Institute, Northeastern University London, UK}

\fi

\hypersetup{
pdftitle={A template for the arxiv style},
pdfsubject={q-bio.NC, q-bio.QM},
pdfauthor={Cosimo Agostinelli, Ivan Casanovas, Lochan Chaudhari, Arda Ergin, Pablo Estévez-Gutiérrez, Akanksha Gupta, Juliane T. Moraes, Mario Edoardo Pandolfo, Carlos Gershenson, Haily Merritt, Andreia Sofia Teixeira},
pdfkeywords={Social Support, Personality Traits, Brain Networks, Mental Health, Complex Systems},
}

\begin{document}
\maketitle

\begin{abstract}
	How are psychosocial profiles, mental health, and brain functional connectivity related? Studies have been dedicated to unraveling the associations of social support perception and neural functional connectivity. Additionally, personality traits have been explored by examining brain networks. Research on mental health has been developed using a broad range of methods and different approaches. However, little attention has been devoted to understanding how personality traits and social variables are related, and to what extent these components are reflected in brain functional connectivity and mental health outcomes. In this work, we aim to address these complex relations by using data from the Human Connectome Project, both from surveys and resting-state fMRI. The survey data includes personality traits measures and self-reported social support-related variables, which we will refer to as inner- and outer-self, respectively. It also includes data on mental health outcomes. Using z-score standardized measures, we analyze correlation matrices to evaluate the association between the inner- and outer-self domains. Our results show that the social indicators are more evidently grouped by impact on social experience than by the duality of inner-outer selves. Using a $k$-means clustering algorithm, we separate individuals into two groups according to social profiles. When confronting these results with the mental health outcomes, we show that the more socially desirable cluster exhibited a higher score on positive aspects such as life satisfaction and purpose in life. In the functional brain connectivity, we observe that the cluster with a more socially beneficial profile exhibits lower interconnectivity, especially in the default mode network. The pipeline we present uses a combined analysis of both fMRI and psychosocial variables, which could open the path for more extensive analysis. 

\end{abstract}


\section{Introduction}

Extensive research has shown that the quality of relationships and social support confers multiple benefits on cognition, behavior, and physiology \citep{Kelly2017, CostaCordella2021, merritt_2025_connection_amp_context, uchino_social_2006}, with social environments further shaping well-being and emotion regulation \citep{beckes_2011_social_baseline_theory}. Such external factors are consequently linked with the functional organization of the brain: for example, socioeconomic status \citep{rakesh_2021_similar_distinct_effects, rakesh_2021_associations_neighborhood_disadvantage} and social network size \citep{Noonan2018, hyon_similarity_2020} predict brain network functional connectivity. More than this, the particular functional connectivity one has represents calibration to environmental experience \cite{merritt_2025_connection_amp_context}. Researchers have coined these contextualized patterns of connectivity \textit{network neuroendophenotypes}. In addition to linking to the environment, the functional organization of the brain also reflects personal and emotional traits relevant to social selves, such as loneliness and meaning in life \citep{Mwilambwe2019, Nostro2018}. Together, these findings suggest that studies trying to understand human behavior and neurophysiology should take into account the social world in which people are embedded \citep{merritt_2025_connection_amp_context}.

\begin{figure}[ht!]
    \centering
    \includegraphics[width=\linewidth]{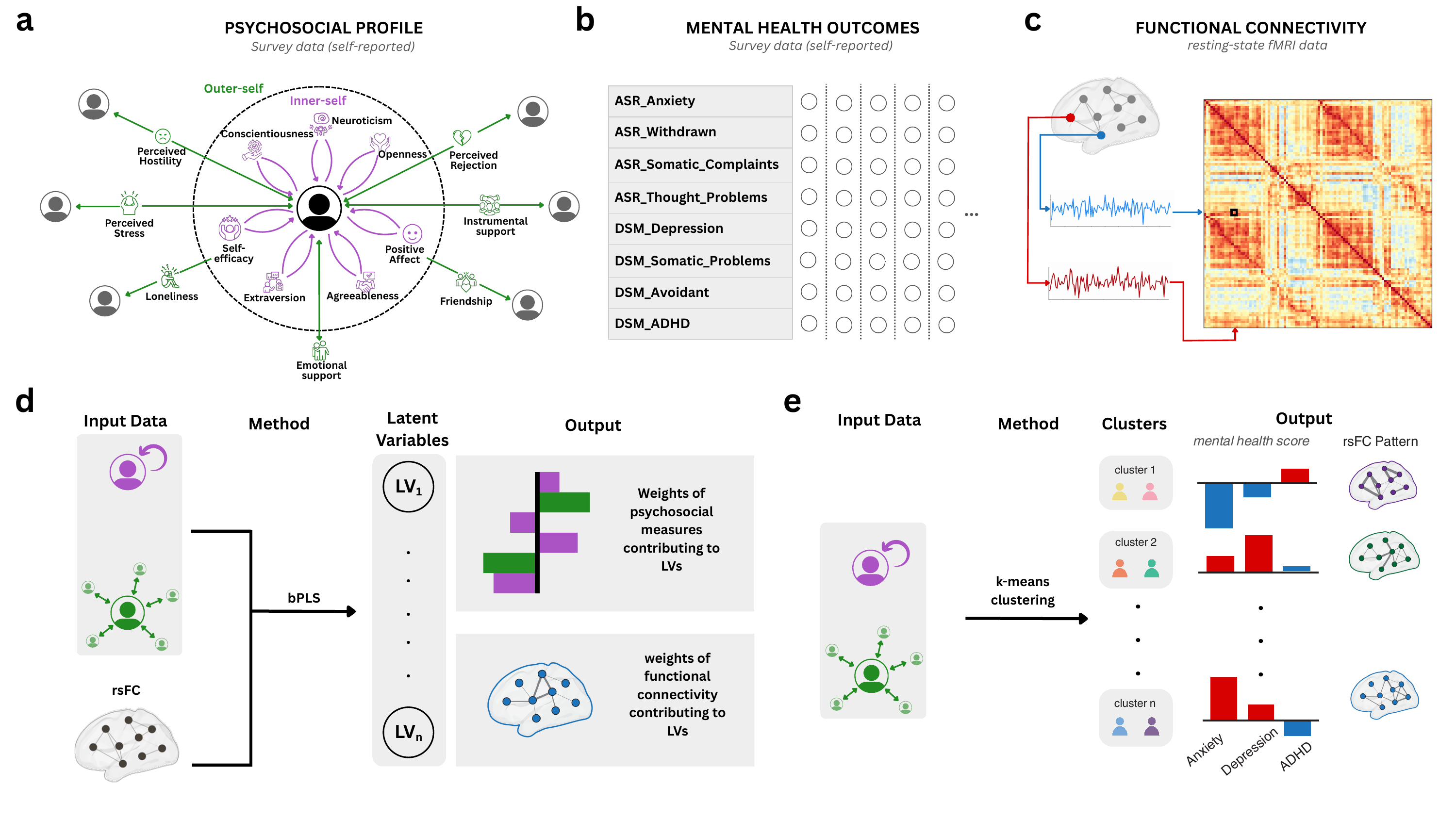}
    \caption{
    \textbf{A schematic of the conceptual framework and analysis pipeline of the study.}
    Our project leverages multimodal data to understand mental health outcomes and their associations with brain organization. In particular, we consider mental health as influenced by personal and contextual features, here represented in (a) with the inner-self (henceforth purple) and outer-self (henceforth green) variables indexed by self-reported survey data. To assess mental health, we use 22 measures from the ASR and DSM (b). We use functional connectivity to capture brain organization (c), which is calculated from the pairwise correlations of resting state fMRI time series across the whole brain, as exemplified by the red and blue nodes. Our analytical workflow proceeds (d) by examining shared associations between psychosocial variables functional connectivity through behavioral Partial Least Squares (PLS). We then characterize how neural associations differ across different psychosocial profiles by first applying \textit{k}-means, then assessing their mental health profiles and corresponding rsFC patterns (e).  
    }
    \label{fig:framework}
\end{figure}

To better understand the relationship between social context and social self, we examined the associations of resting-state functional connectivity (rsFC) with social traits and experiences. We classify these traits as \textit{inner-} and \textit{outer-self} constructs (Figure~\ref{fig:framework}a) using a suite of psychosocial measures. The inner-self refers to characteristics that are relatively intrinsic to the individual and shape how a person engages with their social world, including positive affect, self-efficacy, and personality traits such as agreeableness, openness, conscientiousness, neuroticism, and extraversion. In contrast, the outer-self encompasses aspects of an individual's social environment and interpersonal experiences, including friendship, loneliness, perceived hostility, perceived rejection, emotional and instrumental support, and perceived stress.

Inner- and outer-self constructs can be related to emotional states that are indicators of mental health, such as anxiety and depression (Figure~\ref{fig:framework}b) \cite{Gramstad2013, Penate2020, GabarrellPascuet2023}, and they have been linked to distinct resting state networks (Figure~\ref{fig:framework}c). For instance, the  DMN has been observed to be associated with both types of constructs, though the direction of association is inconsistent \citep{Passamonti2015, Simon2020, MastrandJoergensen2021}. For instance, a higher openness score has been associated with an increased DMN connectivity \citep{Passamonti2015, Simon2020}, but, also a decreased within network DMN connectivity \citep{MastrandJoergensen2021}. Similarly, conscientiousness shows mixed DMN and FPN association \citep{Simon2020}. Loneliness, in turn, upregulates DMN connectivity with limbic, dorsal attention, and somatomotor networks \citep{spreng_default_2020}.

Building on this, functional brain connectivity has been shown to contain fingerprints that could potentially be used to characterize and even to predict cognitive behavior \cite{finn_functional_2015}. A recent study reported that perceived social isolation (loneliness) is associated with functional coupling shifts with increased intra-network connectivity, most expressively in the default mode network (DMN). Also, they show that lonely individuals exhibited increased intra-network coupling for the visual network, and non-lonely ones showed more functionally anti-correlated associations in the DMN to other canonical networks, such as limbic, dorsal attention, and somatomotor \citep{spreng_default_2020}. Another study exploring the patterns of functional connectivity found that inattentive parenting was associated with a lower decrease in connectivity specially between the DMN and the frontoparietal (FPN) and the visual (VIS) networks in children and adolescents \citep{pozzi_investigating_2024}. 

Although multiple studies have investigated associations between resting-state networks and inner- and outer-self constructs, these have typically been examined separately, and to our knowledge, no study to date has jointly considered their shared associations with resting-state networks. To bridge this gap, in this work, we examine associations of these inner- and outer-self constructs with different resting-state networks (Figure~\ref{fig:framework}d). Furthermore, given the link between social support and mental health, we leverage such constructs to group individuals and investigate whether this corresponds to specific mental health  outcomes(Figure~\ref{fig:framework}e). Overall, our work explores the three-way associations among social profiles (inner and outer), mental health outcomes, and brain network connectivity, moving a step further towards a more comprehensive understanding of how the social world shapes the social self.

\section{Methods}
\label{sec:methods}

\subsection{Data description}

We use data from The Human Connectome Project (HCP), a large-scale publicly available resource combining neuroimaging, emotional, personality, social, cognitive, behavioral, and mental health measures in healthy young adults \citep{van2013wu_a}. The full HCP behavioral release includes approximately $N = 1,200$ participants, mainly 22-35 years old. In the present study, we use survey measures available from the broader HCP sample. For analyses with neuroimaging data, however, we focus on \(N = 95\) participants from the Unrelated 100 subset.

Therefore, the dataset comprises two complementary sources of information. First, resting-state fMRI provides an index of intrinsic brain functional connectivity, capturing spontaneous fluctuations in BOLD activity while participants were not engaged in an explicit task \citep{FoxRaichle2007, Biswal2010}. Imaging data were preprocessed using the HCP minimal preprocessing pipeline \citep{glasser2013minimal}, implemented with FSL and FreeSurfer \citep{jenkinson2012fsl, fischl2012freesurfer}, including linear registration and motion correction \citep{jenkinson2002improved}, and also cortical parcellation defined by the Schaefer atlas \citep{schaefer_local-global_2018}. Second, the HCP psychosocial metrics provide a multidimensional description of participants' psychological, social, personality, and mental health profiles. They include measures from the NIH Toolbox Emotion and Social Relationship instruments \citep{Salsman2013, Cyranowski2013}, the NEO Five-Factor Inventory \citep{McCraeCosta2004}, and the Achenbach Adult Self-Report \citep{Achenbach2003}. All survey variables were \(z\)-scored before analysis, allowing measures originating from different instruments and scoring systems to be placed on a common scale. Together, these measures enable the characterization of individual differences in subjective well-being, perceived social relationships, personality traits, and psychiatric symptom burden in relation to intrinsic brain organization \citep{Mwilambwe2019, merritt2024}.


Psychosocial variables were organized into two theoretically motivated domains. The first domain we label \textit{inner-self} measures, which capture properties of the individual's own personality or capacities. The measures included in this domain are: \textit{Positive Affect}, \textit{Self-efficacy}, and the Big Five personality traits: \textit{Openness}, \textit{Conscientiousness}, \textit{Extraversion}, \textit{Agreeableness}, and \textit{Neuroticism}. The second domain we label \textit{outer-self} measures and defines as an individual's perceptions of the social and relational environment. Specific indices of this domain included \textit{Emotional Support}, \textit{Instrumental support}, \textit{Friendship}, \textit{Loneliness}, \textit{Perceived Rejection}, and \textit{Perceived Hostility}. 

Additionally, we assess a suite of \textit{mental health outcomes} from two difference scales, the Achenbach Adult Self-Report (which included syndrome scales and broadband indices) and DSM symptom dimensions. Together, they indexed internalizing symptoms, externalizing symptoms, somatic complaints, attention problems, thought problems, aggressive and rule-breaking behavior, depressive and anxiety problems, avoidant personality problems, attention-deficit/hyperactivity-related problems, and antisocial personality problems.

\subsection{Partial Least Squares analysis}


To examine shared multivariate associations between intrinsic functional connectivity and the psychosocial measures, we use behavioral Partial Least Squares (bPLS) \citep{McIntosh1996, McIntoshLobaugh2004, Krishnan2011}. This multivariate, data-driven method simultaneously maximizes the covariance between two data matrices and decomposes this shared covariance into a set of orthogonal latent variables (LVs). This approach is well-suited to our data because they are intrinsically high-dimensional and multi-collinear: resting-state connectivity comprises many correlated network features, whereas the behavioral data comprised multiple interrelated measures spanning inner and outer-self variables, and mental health outcomes.

Let $\mathbf{X}$ denote the $N \times P$ matrix of z-scored functional connectivity features for $N$ participants and $P$ brain features, and let $\mathbf{Y}$ denote the $N \times Q$ matrix of z-scored behavioral measures for the same participants and $Q$ survey variables. For a specific domain, the brain--behaviour cross-covariance matrix was computed as:
\begin{equation}
    \mathbf{R} = \frac{1}{N-1} \mathbf{X}^{T}\mathbf{Y} .
\end{equation}
Then, singular value decomposition (SVD) was applied to $\mathbf{R}$:
\begin{equation}
    \mathbf{R} = \mathbf{U} \mathbf{\Sigma} \mathbf{V^{T}},
\end{equation}
where $\mathbf{U}$ ($P \times K$) and $\mathbf{V}$ ($Q \times K$) are orthogonal matrices containing the left and right singular vectors, respectively, and $\Sigma$ is a diagonal matrix of singular values. The singular vectors $U$ and $V$ represent the brain and behavioral loadings, indicating the relative contribution of each original feature to the multivariate pattern. The singular values in $\Sigma$ indicate the amount of cross-covariance explained by each corresponding LV. 
Participant-specific brain and behavior scores are computed by projecting the original data onto their respective loadings:
\begin{equation}
    \mathbf{L_X} = \mathbf{X} \mathbf{U}, \quad \mathbf{L_Y} = \mathbf{Y} \mathbf{V}
\end{equation}
and the correlation between $\mathbf{L_X}$ and $\mathbf{L_Y}$ is used to characterize the strength and direction of the multivariate brain-behavior association for each LV.


The statistical significance of each LV is assessed via permutation testing\citep{Krishnan2011,Mwilambwe2019}. On each of $2,000$ iterations, the rows of $\mathbf{Y}$ were randomly permuted, destroying the participant-level brain-behavior correspondence while preserving the marginal distributions of both matrices. The cross-covariance matrix is recomputed and a new set of singular values extracted, generating an empirical null distribution of singular values under the hypothesis of no brain--behaviour covariance.
The $p$-value for each LV is defined as the proportion of permuted singular values that equalled or exceeded the observed value. Only LVs with $p < 0.05$ are considered statistically significant and interpreted in later analysis.

For each significant LV, the reliability of individual loading elements is assessed via nonparametric bootstrap resampling \citep{Efron1993,Krishnan2011}. On each of $2,000$ iterations, participants are resampled with replacement, bPLS is rerun, and new salience vectors are obtained. We then compute a bootstrap ratio (BSR) for each loading element $i$ on a specific LV. Only absolute value of the BSR$>2$ are considered reliable \citep{Mwilambwe2019}.

\subsection{Cluster Analysis}

The $k$-means algorithm partitions $n$ observations into $k$ mutually exclusive
clusters such that each observation is assigned to the cluster whose centroid is
nearest in feature space, minimising the within-cluster sum of squares (WCSS) \citep{lloyd_least_1982, ikotun_k-means_2023}.
The optimal number of clusters $k$ was selected by jointly evaluating four complementary validity indices: the WCSS elbow criterion, the Silhouette Score \citep{rousseeuw_silhouettes_1987}, the Calin\-ski--Harab\'{a}sz index \citep{calinski_dendrite_1974}, and the Davies--Bouldin index \citep{davies_cluster_1979}. To assess whether the derived cluster solution produced statistically significant differences in the variables of interest, we adapt inferential tests to the number of clusters identified. For $k = 2$, group differences are evaluated with an independent-samples Student's $t$-test \citep{student_probable_1908}.
For $k > 2$, a one-way analysis of variance (ANOVA) is employed \citep{fisher_statistical_1925}, followed by post-hoc pairwise comparisons with Bonferroni correction to control the family-wise error rate. A significance threshold of $\alpha = 0.05$ is adopted throughout.

Code for all analyses is available at \href{https://github.com/h-merritt/complexity72h.git}{https://github.com/h-merritt/complexity72h.git}.

\section{Results and Discussion}
\label{sec:results-discussion}

\subsection{Contrasting Theory-Driven and Data-Driven Approaches to Grouping Psychosocial Measures}

\begin{figure}[t!]
    \centering
\includegraphics[width=1.1\textwidth]{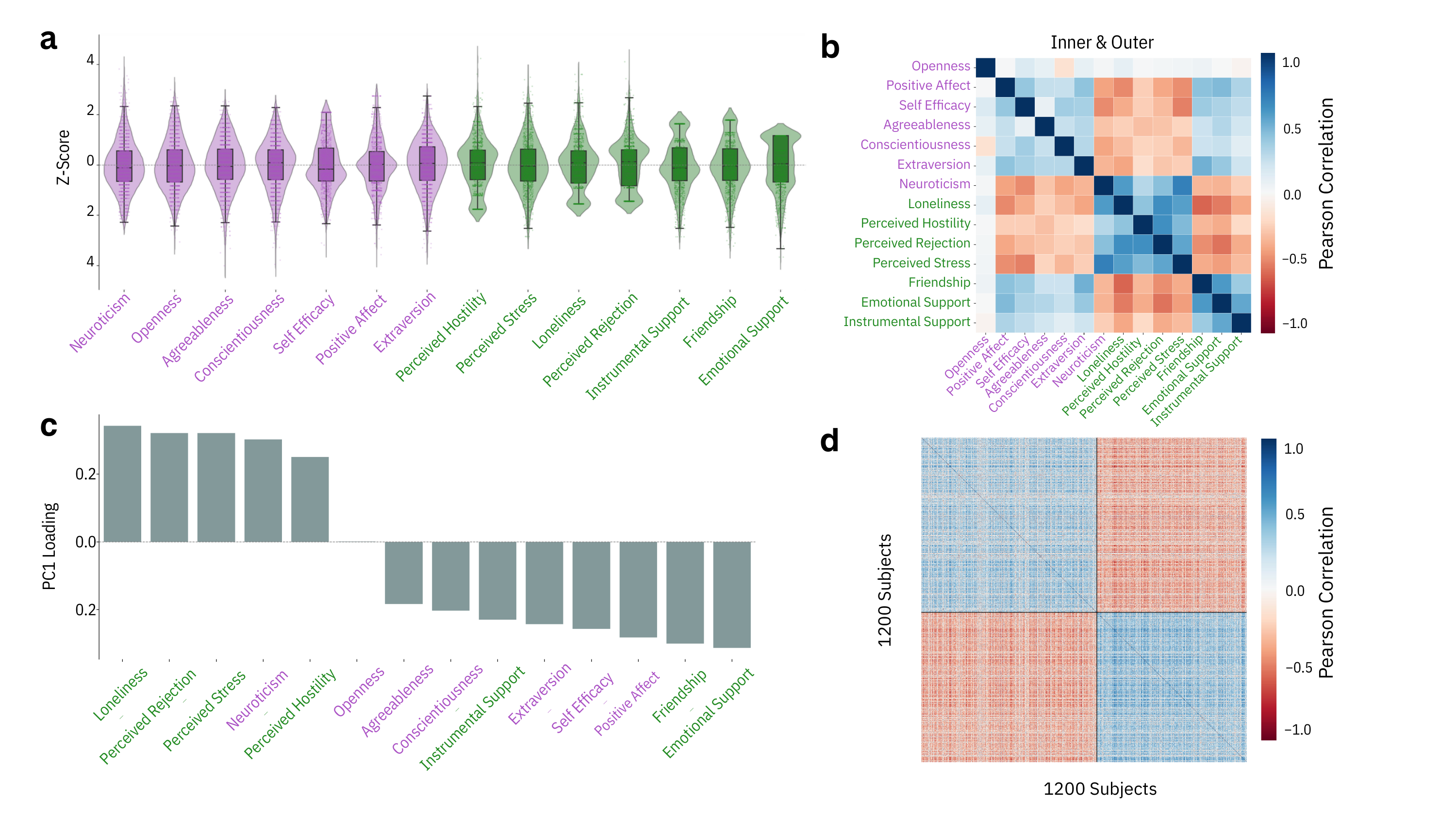}
    \caption{\textbf{Characterization of psychosocial measures.}
(a) Distributions of standardized inner-self (purple) and outer-self (green) measures expressed as z-scores.
(b) Cross-correlation matrices of all psychosocial measures, separated by inner-self (purple) and outer-self (green).
(c) Loadings of psychosocial measures on the first principal component. 
(d) Pairwise similarity between subjects according to their scores on psychosocial measures.}
    \label{fig:2}
\end{figure}

We first explore the descriptive and correlational structure of psychosocial measures to characterize how inner- and outer-self constructs are organized before assessing their relationship with brain connectivity and mental health outcomes. Since these constructs use different scales, we $Z$-scored all measures so they can be analyzed unambiguously in the same space. Distributions of all psychosocial measures can be see in Figure~\ref{fig:2}a, in which we observe bimodality for some measures (all of which fall into our \textit{Outer} label). 

We next examine the relationship between all the psychosocial measures to assess the extent to which the inner-outer distinction is borne out. In Figure~\ref{fig:2}b, we observe that psychosocial measures are most dominantly grouped by valence (i.e., \textit{Perceived Hostility} and \textit{Perceived Rejection} are positively correlated with each other and negatively correlated with \textit{Friendship}) rather than by the dichotomy inner-outer selves. \textit{Openness} stands out as the most independent score, not being aligned with either of the two groups. If we look at the correlation considering the inner and outer labels (Figure~\ref{fig:2}b), outer labels are more strongly divided by valence, while inner variables seem to be more useful to describe high-sociability, as they are all involved in said cluster except for \textit{Neuroticism} and \textit{Openness}. This may also be appreciated when comparing the Pearson correlation of psychosocial scores, (Figure~\ref{fig:2}d) where subjects are visibily grouped in two similar clusters.

To verify that the data-driven grouping identified in the correlation structure is preserved under dimensionality reduction, we perform PCA on the full set of variables. The cumulative explained variance reveals that seven principal components (PCs) are required to account for $80\%$ of the total variance (Figure~\ref{fig:acc_variance_all_pca}a, Appendix~\ref{sec:appendixA}), yet PC1 alone captures $40\%$, indicating a dominant axis of variation in the data. 
Crucially, this first component recapitulates the distinction in valence observed in the correlation matrix: PC1 loadings segregate systematically according to sociability group, with signs aligning across variables within each group (Figure~\ref{fig:2}c).

None of the remaining PCs exhibits a systematic preference for inner- over outer-self constructs, indicating that the presented dichotomy does not emerge as a latent structure in the combined variable set.
A notable exception concerns \textit{Openness}, which did not conform to the grouping in the correlation matrix and emerges here as the dominant contributor to both PC2 and PC3 (Figure~\ref{fig:combined_pca_top_loadings}b,c, Appendix~\ref{sec:appendixA}). This consistent prominence across components suggests that \textit{Openness} captures a dimension of individual variability that may be orthogonal to sociability, warranting further investigation in future work. 
We also perform PCA for inner- and outer-self constructs to look for any internal structure within each group. Their accumulated explained variance suggests a much simpler structure than the combined set of variables: outer variables require $2$ PCs to get to $70\%$ variance explained, inner variables require $3$ PCs (Figure~\ref{fig:acc_variance_all_pca}b,c, Appendix~\ref{sec:appendixA}). If we now focus on the loadings of the two first PCs for each subset of measures, results concur with the above findings: signs align with the sociability groups; and a strong influence of the \textit{Openness} score, in this case only for the inner scores (Figure~\ref{fig:pca_inner_outer_top_loadings} in Appendix~\ref{sec:appendixA}).

\subsection{Shared associations of psychosocial measures with functional connectivity}
\begin{figure}[t]
    \centering
    \includegraphics[width=0.7\textwidth]{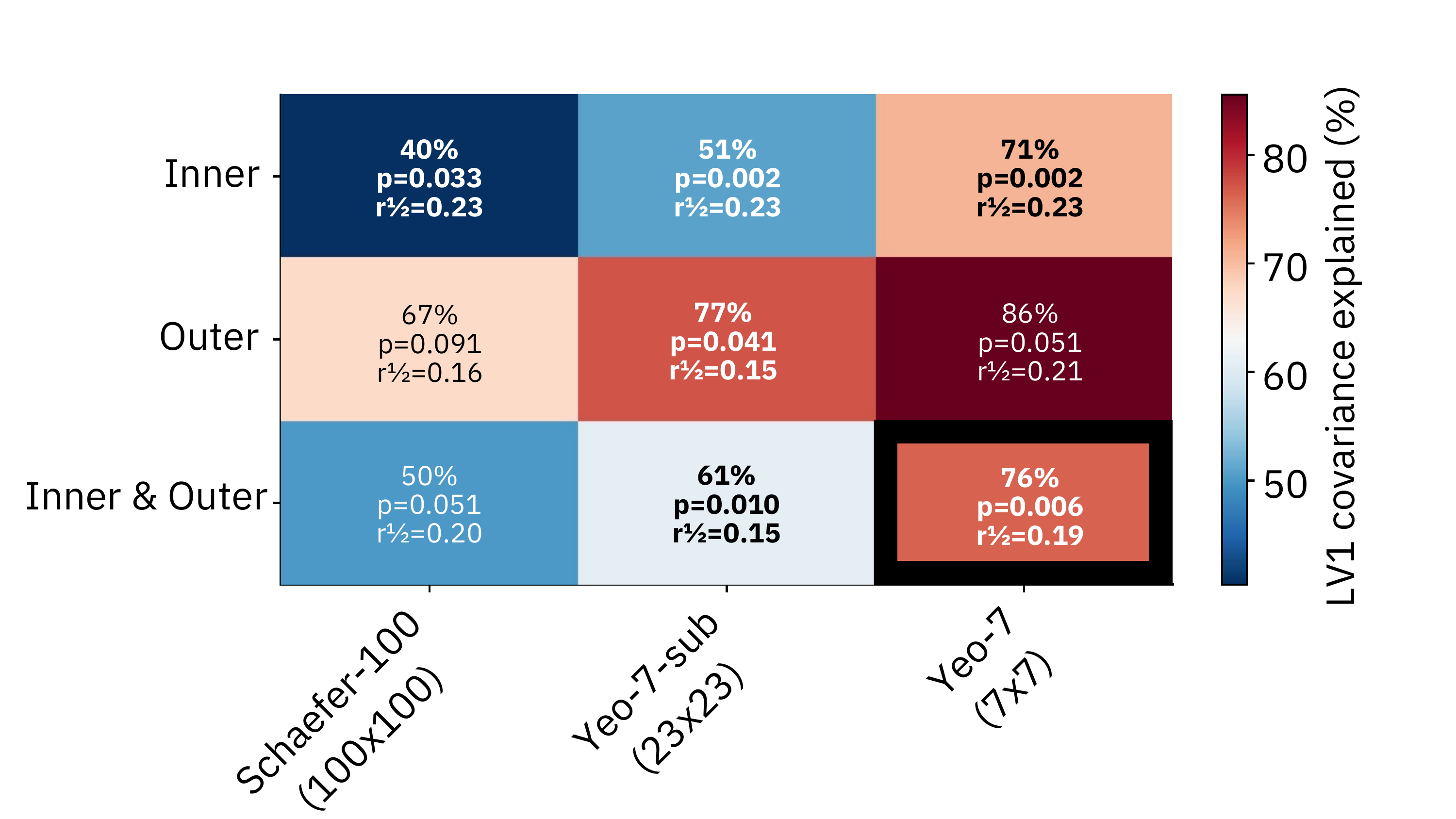}
    \caption{\textbf{Overview of PLS Analyses.} Color indicates the percentage of LV1 covariance explained across resolutions of functional connectivity (x-axis) and sets of psychosocial measures (y-axis). Each PLS analysis yielded at most one significant latent variable (LV1), with $p$-values determined by permutation tests (variables with $p$-values $<0.05$ are bold highlighted). }
    \label{fig:pls}
\end{figure}

Of the nine PLS analysis we ran, we find that the first latent variables from coarser parcellations of the functional connectivity matrices and the outer measures explain more of the variance (Figure~\ref{fig:pls}). In particular, the analysis whose first latent variable explained the most variance employed only the outer measures and included connectivity averaged within and between the Yeo-7 systems \citep{yeo2011organization}, with 86\% variance explained ($p$ = 0.051; see Figure~\ref{fig:pls}). Given that we are interested in shared variance between \textit{all} psychosocial measures and functional connectivity, we also investigate further the PLS analyses using all 14 psychosocial measures. Of the three resolutions with these data, the Yeo-7 scale explains the most variance (76\%, $p=0.006$), so we focus the remainder of our discussion on this latent variable. 

\begin{figure}[t]
    \centering 
    \includegraphics[width=\textwidth]{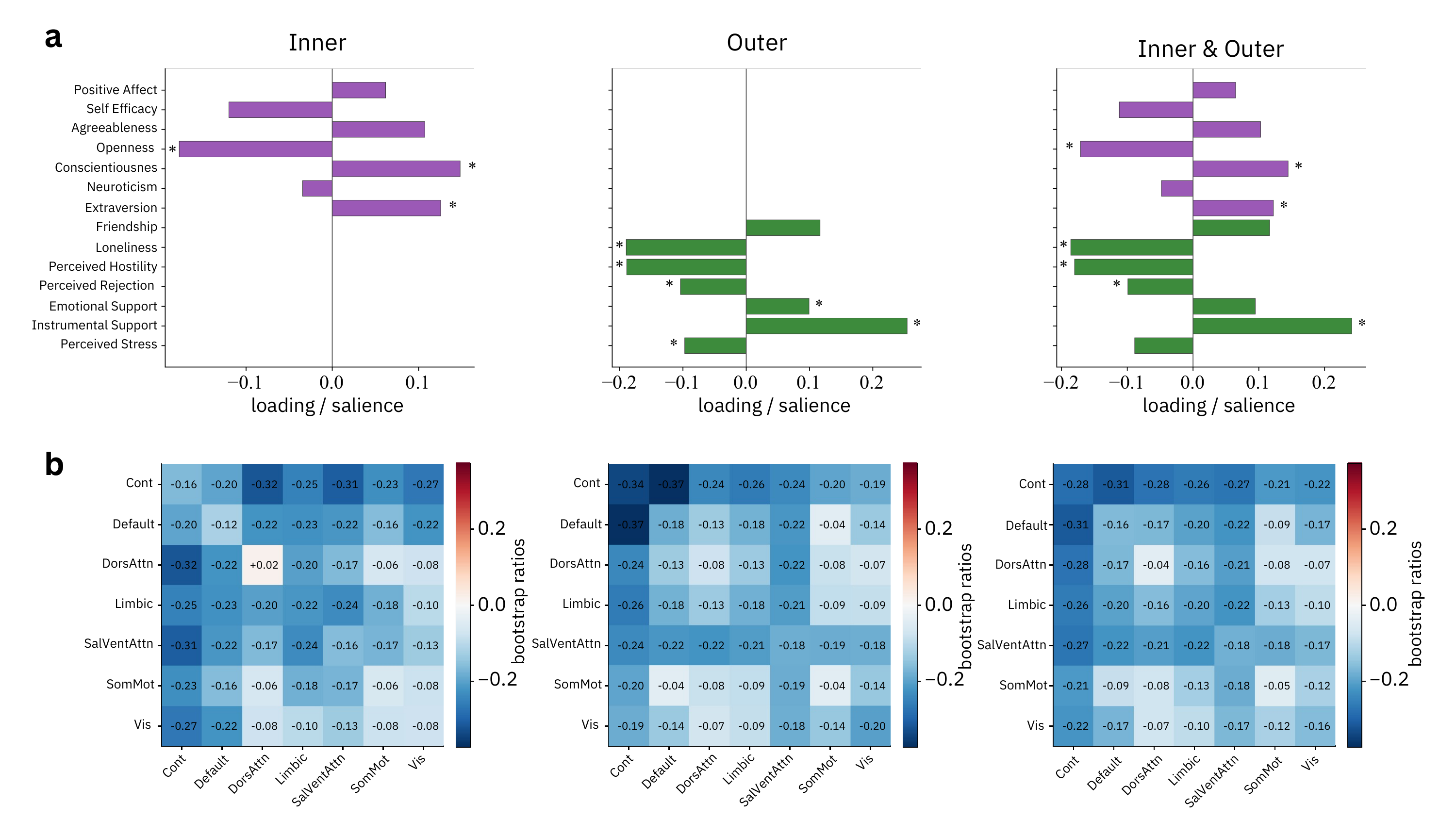}
    \caption{\textbf{Behavioral Partial Least Squares analysis of associations between psychosocial measures and the coarsest resolution of functional connectivity.} The leftmost column corresponds to the analysis using only inner-self measures, the center column corresponds to the analysis using only outer-self measures, and the rightmost column corresponds to the analysis using all psychosocial measures. (a) Loadings on the first latent variable of each psychosocial measure included in the analysis. Asterisks indicate measures that load reliably based on bootstrap ratios. (b) Color coded bootstrap ratios of the loadings of systemwise functional connectivity on the first latent variable of each analysis. The colorbar depicts the value of the bootstrap ratio.
    }
    \label{fig:pls_full_grid}
\end{figure}

In Figure~\ref{fig:pls_full_grid}a, we examine the loadings of the behavioral measures on this first latent variable, we find that \textit{Instrumental Support} loads strongly and positively. \textit{Loneliness, Perceived Hostility, Openness} and \textit{Perceived Rejection} load negatively, while \textit{Conscientiousness} and \textit{Extraversion} exhibit small positive loadings. With respect to the functional connectivity (Figure~\ref{fig:pls_full_grid}b), negative loadings are distributed across the entire brain. Put together, these results tell us that lower connectivity is associated with more \textit{Instrumental Support, Conscientiousness,} and \textit{Extraversion} and less \textit{Loneliness, Perceived Hostility, Openness,} and \textit{Perceived Rejection}, in line with \cite{mwilambwe2023age}. 

\subsection{Divergent functional connectivity across distinct psychosocial profiles}

\begin{figure}[ht!]
    \centering
    \includegraphics[width=0.90\textwidth]{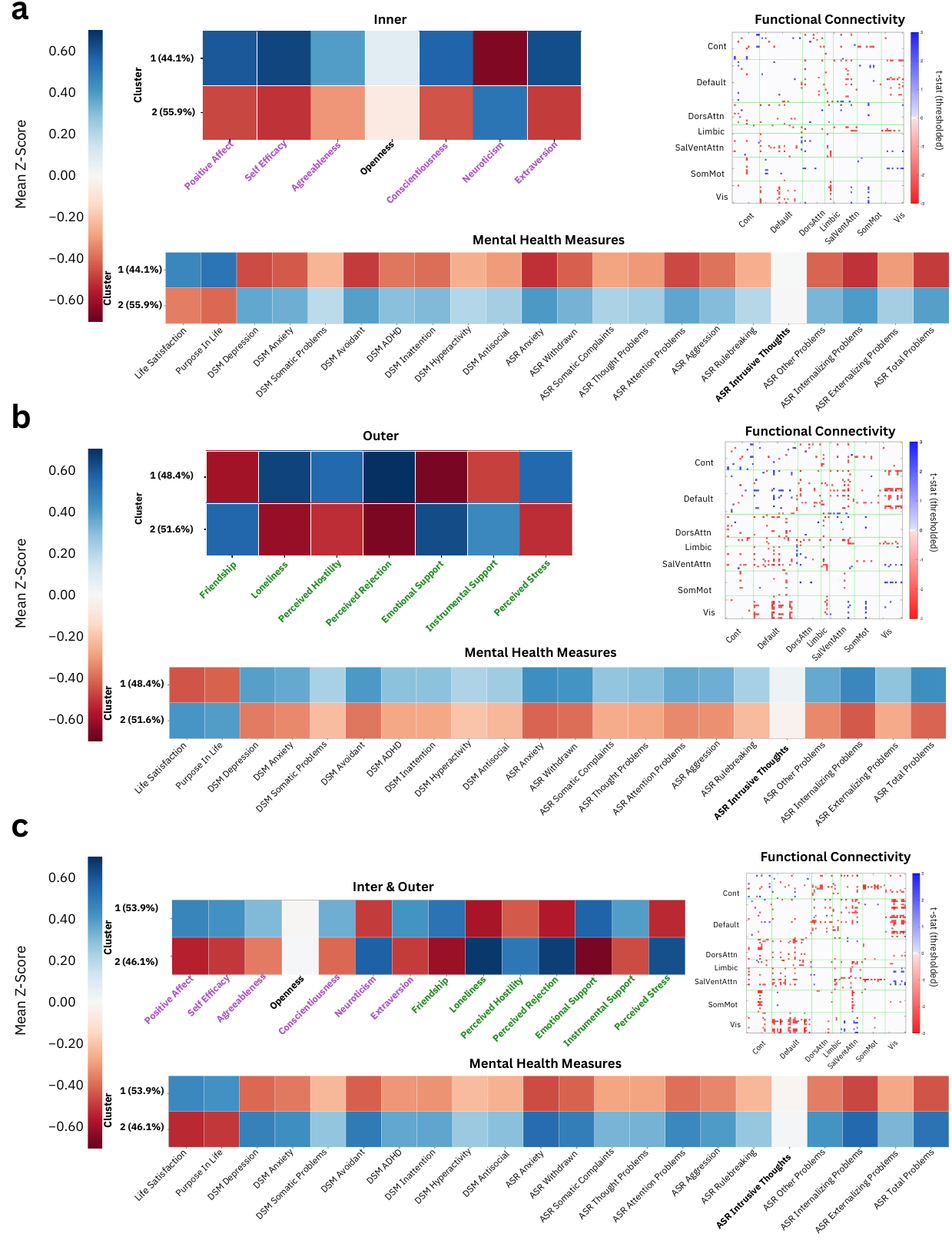}
    \caption{ \textbf{Psychosocial profiles and their distinct mental health outcomes and functional connectivity patterns.} All profiles were identified with $k=2$ using either the inner-self measures (a), outer-self measures (b), or all psychosocial measures (c). Within each panel, the top figure depicts cluster profiles, the bottom figure shows mental health outcomes, and the right figure indicates where there were differences in functional connectivity (thresholded to show only significant differences).  
    }
    \label{fig5}
\end{figure}

We assessed the fit statistics for $k$ ranging from 1-9 applied to all measures, only the inner measures, and only the outer measures. For the clustering of all measures and of only inner measures, $k=2$ best described the clusters (Figure \ref{fig:cluster-metrics} , Appendix \ref{sec:appendixB}), so we compare partitions at this resolution. When all psychosocial measures were used to cluster individuals, the clusters of the resulting partition are characterized by their valence: either more social desirable or socially less desirable. Specifically, cluster 1, which contains 53.9\% of individuals, has higher \textit{Positive Affect, Friendship,} and \textit{Emotional Support} and lower \textit{Neuroticism, Loneliness,} and \textit{Perceived Stress}; cluster 2 is the perfect complement, with the two clusters differing significantly on all measures (all $|t|$s > 12.43, all $p$s $6.8 \times 10^{-32}$) except \textit{Openness} ($t = -0.35, p = 1$). We investigated the mental health outcomes of these two clusters, and found that the more socially desirable cluster scored higher in \textit{Life Satisfaction} and \textit{Purpose in Life} and expressed better mental health for every measure. Again, the less socially desirable cluster represented the complement, differing significantly on all measures (all $|t|$s > 8.8, all $p$s $< 2.46 \times 10^{-15}$) except \textit{Intrusive Thoughts} ($t=-0.42, p = 1$). Upon examining the differences in functional connectivity between the clusters, we find that the more socially desirable cluster in general had lower connectivity, but especially between Default and each of Visual, Salience Ventral Attention, and Dorsal Attention systems and also between Dorsal Attention and Control systems. The less socially desirable cluster, on the other hand, only exhibited higher connectivity between Visual and Salience Ventral Attention systems (see Figure\ref{fig5}c). 

The results with just the inner and just the outer measures follow a similar narrative, with one more socially desirable cluster that had better mental health and one less socially desirable cluster (comprising 44.1\% of people for inner measures and 51.6\% of people for outer measures) that had worse mental health. Specifically, when we clustered using only the inner-self measures, the two clusters differed significantly on all psychosocial measures (all $|t|$s $>12.7$, all $p$s $10 \times 33^{-10}$) except \textit{Openness} ($t = 2,43, p = 0.44$) and on all mental health measures (all $|t|$s > 7.7, all $p$s < 4.78) except \textit{Intrusive Thoughts} ($t = 0.15, p = 1$) (see Figure\ref{fig5}a). Similarly, when we clustered using only the outer-self measures, the two clusters differed significantly on all measures (all $|t|$s > 18.26, all $p$s $2.37 \times 10^{-7}$) (see Figure\ref{fig5}b). Differences in functional connectivity were more sparse when the clusters were defined by only the inner measures, but they tended to occur between Default and each of Visual, Salience Ventral Attention, and Dorsal Attention systems and in the same direction as when all measures were used. Clusters defined by only outer measures had more pronounced differences in functional connectivity and diverged from the above patterns only in that the less socially desirable cluster exhibited higher connectivity within the Default Mode Network (DMN).

\section{Conclusions}
\label{sec:conclusion}

Our analysis revealed that psychosocial measures covary mainly with respect to the valence of social experience than around the inner/outer distinction itself. PLS revealed that the shared covariance of inner and out psychosocial measures was linked with lower whole-brain functional connectivity at a coarse resolution. When we assess differences in psychosocial profiles using the inner and outer measures, we find they are complementary, splitting according to valence. The more socially desirable cluster in each analysis exhibits better mental health outcomes and differences in whole-brain functional connectivity. Taken together, these results support the idea that inner- and outer-self constructs, mental health outcomes, and rsFC form an integrated system.

A relevant limitation of the present work is that the subset of HCP participants in our analysis are generally healthy adults \citep{van2013wu_a}, constraining our ability to make wide-sweeping claims about psychopathology. Thus, in future work, this framework could also be applied to  with populations with anxiety disorders, depression, or social disorders to test whether the psychosocial profiles and their neural correlates are diagnostically informative. Additionally, if psychosocial social profiles predict mental health outcomes, interventions (e.g., social support programs) could be evaluated for corresponding changes in connectivity.

Another limitation of our study is its cross-sectional design: we relate a static psychosocial profile to a static connectivity pattern, captured at a single time-point, but human social environments are dynamic~\citep{gonzalez2025evidence}. It is therefore unclear the extent to which the associations we identify are themselves static or dynamic. Testing this would require longitudinal data linking the dynamics of social environment change to functional brain network reconfiguration — a natural extension of the present work.

Ultimately, building a unified framework capturing social context, brain functional connectivity and mental health outcomes allows for a comprehensive understanding the diversity of how humans navigate their social world and any consequences associated with this. This may be prove essential for understanding not only the healthy social brain, but also mechanisms underlying mental health disorders.

\section{Acknowledgements}
This work is the output of the workshop Complexity72h by Complexity Next Gen, held at Northeastern University London, London, UK, 22-26 June 2026. www.complexitynextgen.org/complexity72h/.
C.A. acknowledges support by the “BeyondTheEdge: Higher-Order Networks and
Dynamics” project (European Union, REA Grant Agreement No. 101120085).

\appendix
\section{PCA results}
\label{sec:appendixA}


As the 'inner' and 'outer' are \textit{a priori} labels, we also apply dimensionality reduction to the psychosocial measures to assess the extent to which this distinction is supported by the dominant modes of variability in the data, or on the contrary which latent variables are most relevant. Specifically, we use Principal Component Analysis (PCA), which is a linear dimensionality reduction tool that, given a certain standardised dataset $\mathbf{Z}$ of $N$ correlated variables $X_1,...,X_N$, linearly combines them into principal components (PCs) orthogonal to each other \citep{Fleming2023, ShamsEldin2025}. Within each PC we may order the most relevant variables according to how much variance they explained. As we project the data onto the eigenvectors of the correlation matrix, we are able to understand better the underlying structure of our dataset.

We build the correlation matrix following the equation:
\begin{equation}
    \text{Corr}(\mathbf{Z}) = \frac{1}{N-1}\mathbf{Z}^T\mathbf{Z}
    \label{eq:corr}
\end{equation}
We obtain the normalised eigenvectors $(PC_1,...,PC_N$) and eigenvalues $(\lambda_1,...,\lambda_N)$ so that $|\lambda_1|>|\lambda_2|>...>\lambda_N$. By looking at the components of each $PC$ (loadings) we will have a measure of the relevance of each variable. We can also use eigenvalues to compute how much explained variance is associated to each PC:
\begin{equation}
    \text{EV}_i = \frac{\lambda_i}{\sum^N_j \lambda_j}
\label{eq:EV}
\end{equation}

By leveraging both theory- and data-driven techniques in this way, we can compare how each represents the data. 

\begin{figure}[ht!]
    \centering

    \begin{subfigure}{0.32\linewidth}
        \centering
        \includegraphics[width=\linewidth]{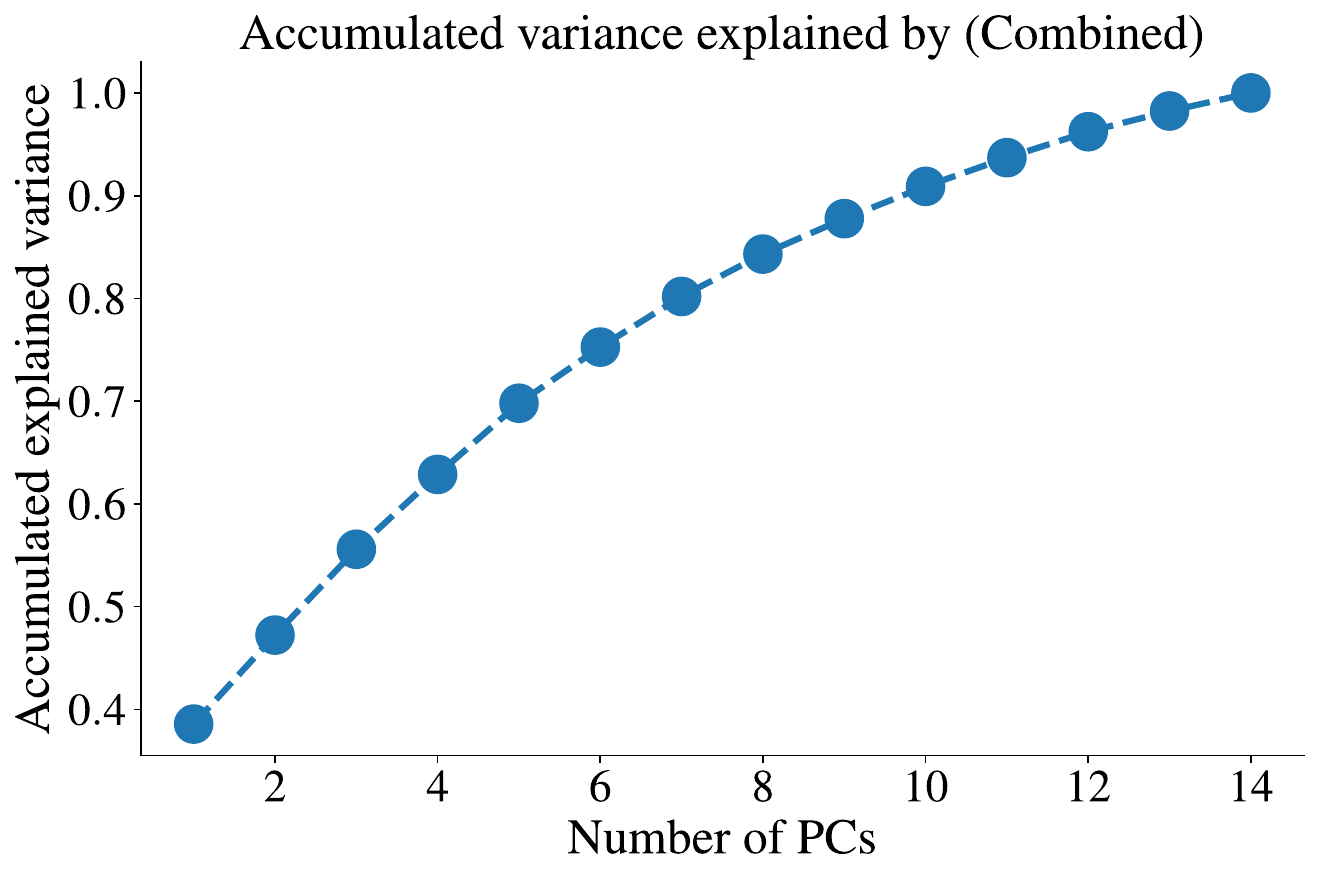}
        \caption{Combined PCA}
        \label{fig:acc_variance_combined_pca}
    \end{subfigure}
    \begin{subfigure}{0.32\linewidth}
        \centering
        \includegraphics[width=\linewidth]{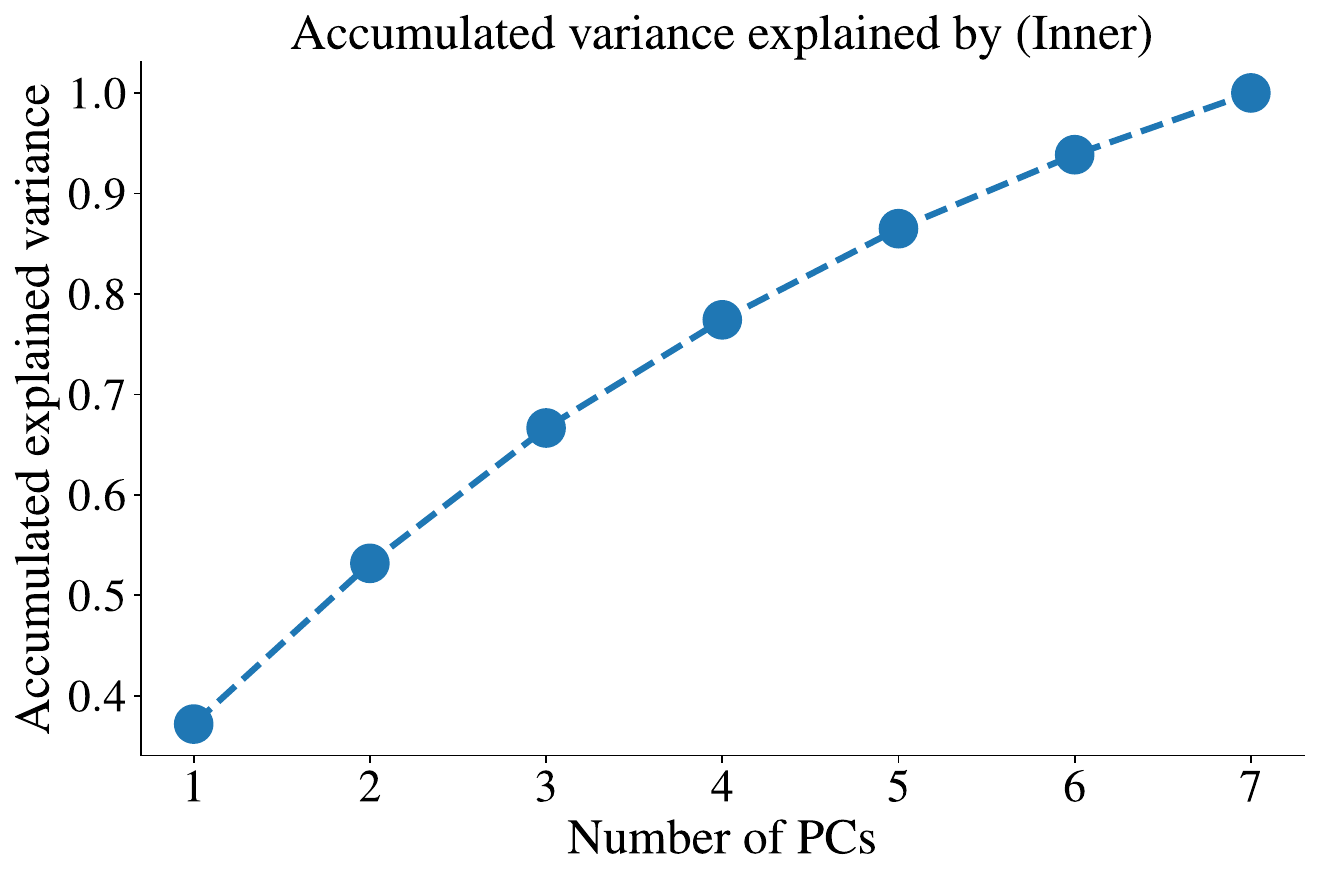}
        \caption{Inner PCA}
        \label{fig:acc_variance_inner_pca}
    \end{subfigure}
    \begin{subfigure}{0.32\linewidth}
        \centering
        \includegraphics[width=\linewidth]{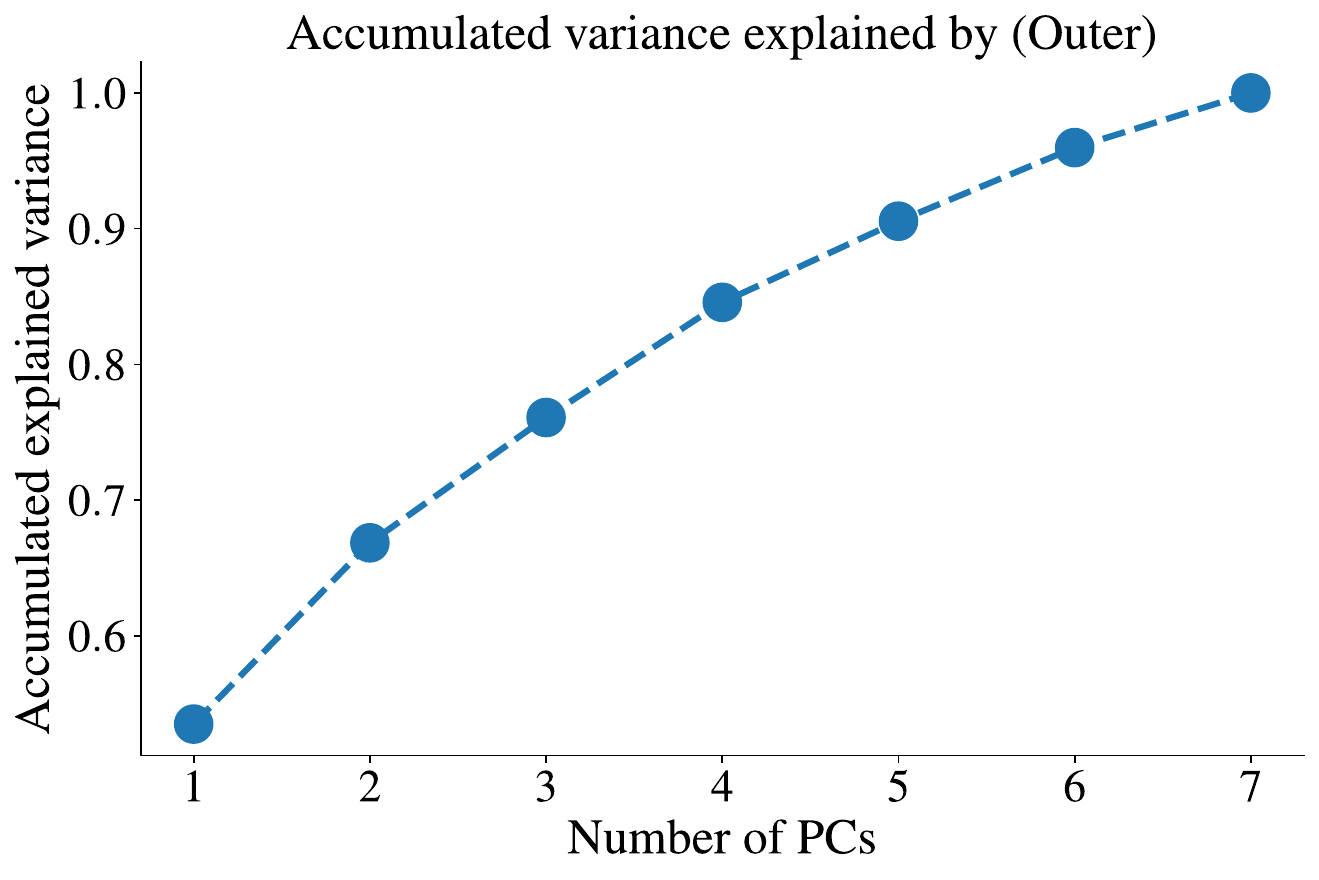}
        \caption{Outer PCA}
        \label{fig:acc_variance_outer_pca}
    \end{subfigure}

    \caption{Accumulated explained variance by number of principal components considered for combined (a), inner (b), and outer (c) PCA analyses.}
    \label{fig:acc_variance_all_pca}
\end{figure}

\begin{figure}[ht!]
    \centering

    \begin{subfigure}{0.32\linewidth}
        \centering
        \includegraphics[width=\linewidth]{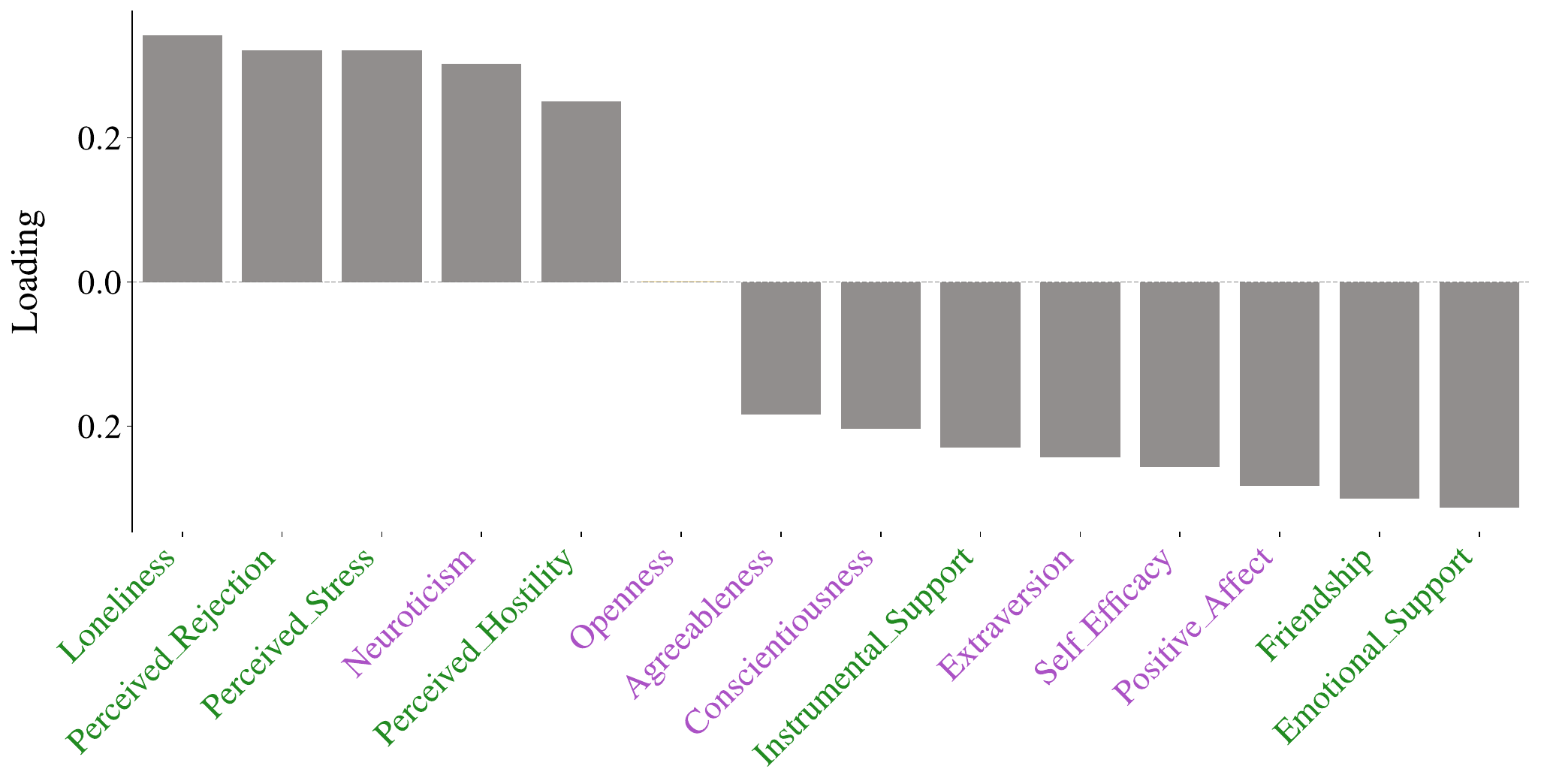}
        \caption{PC1}
        \label{fig:pc1}
    \end{subfigure}
    \begin{subfigure}{0.32\linewidth}
        \centering
        \includegraphics[width=\linewidth]{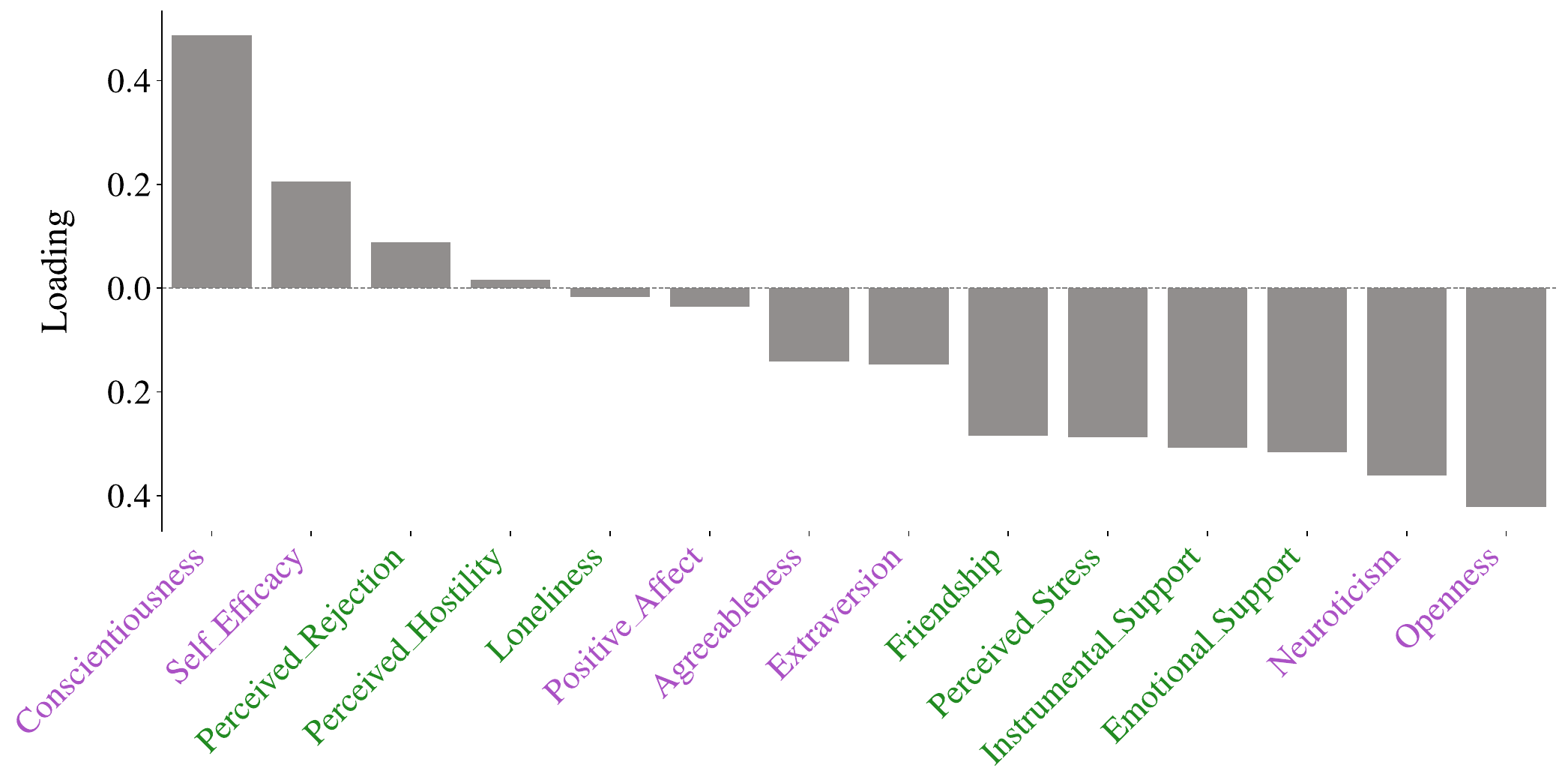}
        \caption{PC2}
        \label{fig:pc2}
    \end{subfigure}
    \begin{subfigure}{0.32\linewidth}
        \centering
        \includegraphics[width=\linewidth]{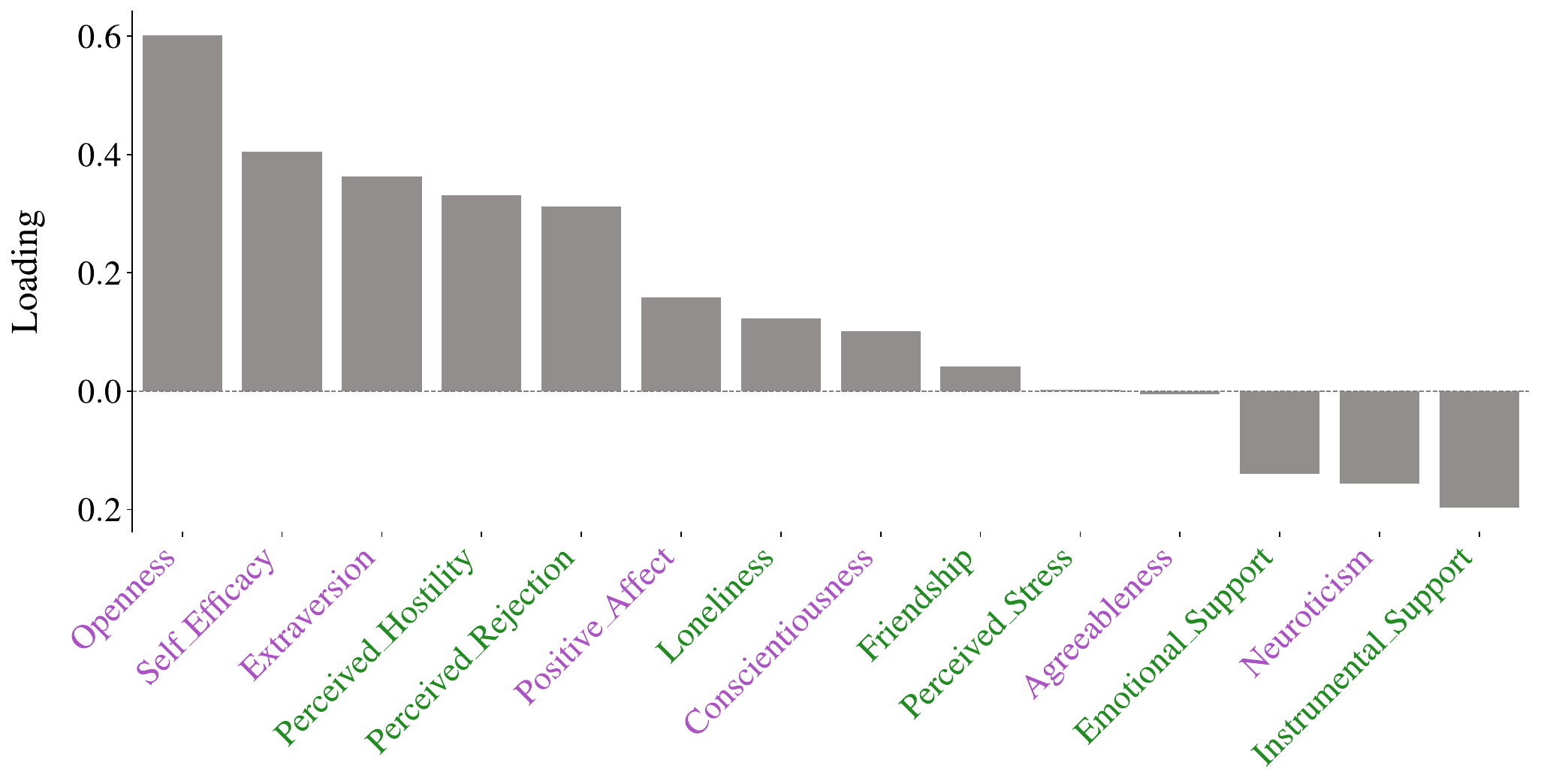}
        \caption{PC3}
        \label{fig:pc3}
    \end{subfigure}

    \vspace{0.4cm}

    \begin{subfigure}{0.32\linewidth}
        \centering
        \includegraphics[width=\linewidth]{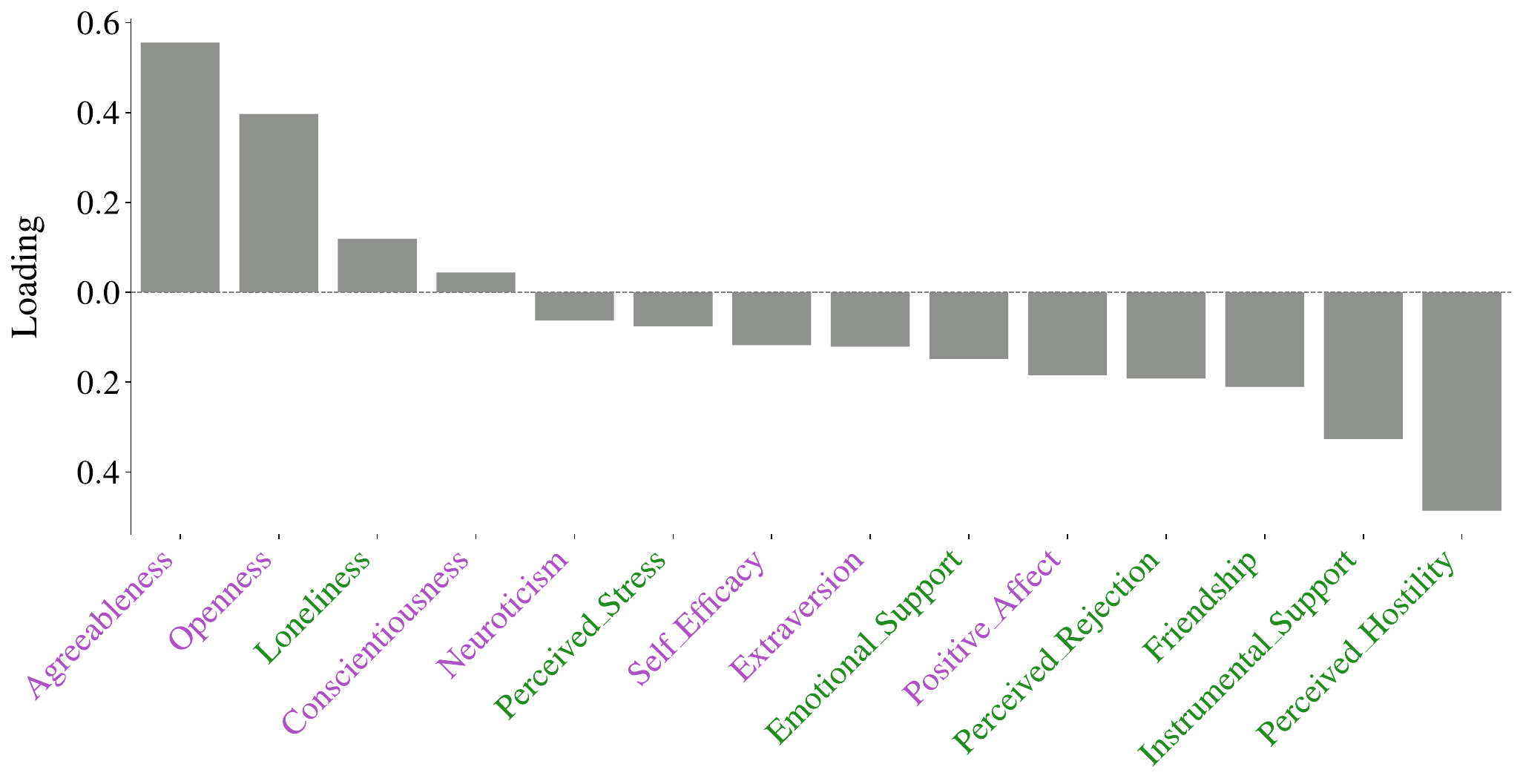}
        \caption{PC4}
        \label{fig:pc4}
    \end{subfigure}
    \begin{subfigure}{0.32\linewidth}
        \centering
        \includegraphics[width=\linewidth]{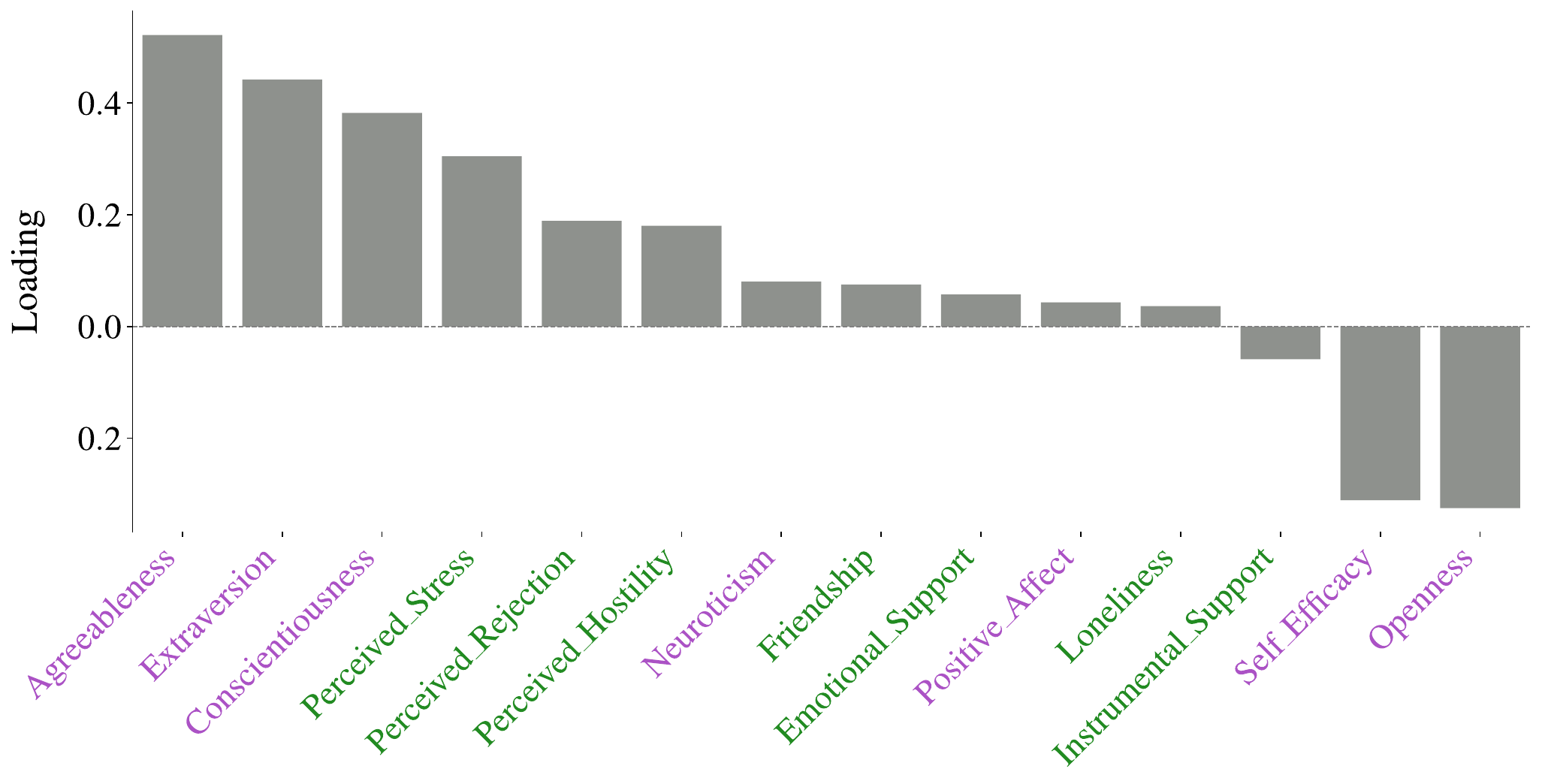}
        \caption{PC5}
        \label{fig:pc5}
    \end{subfigure}
    \begin{subfigure}{0.32\linewidth}
        \centering
        \includegraphics[width=\linewidth]{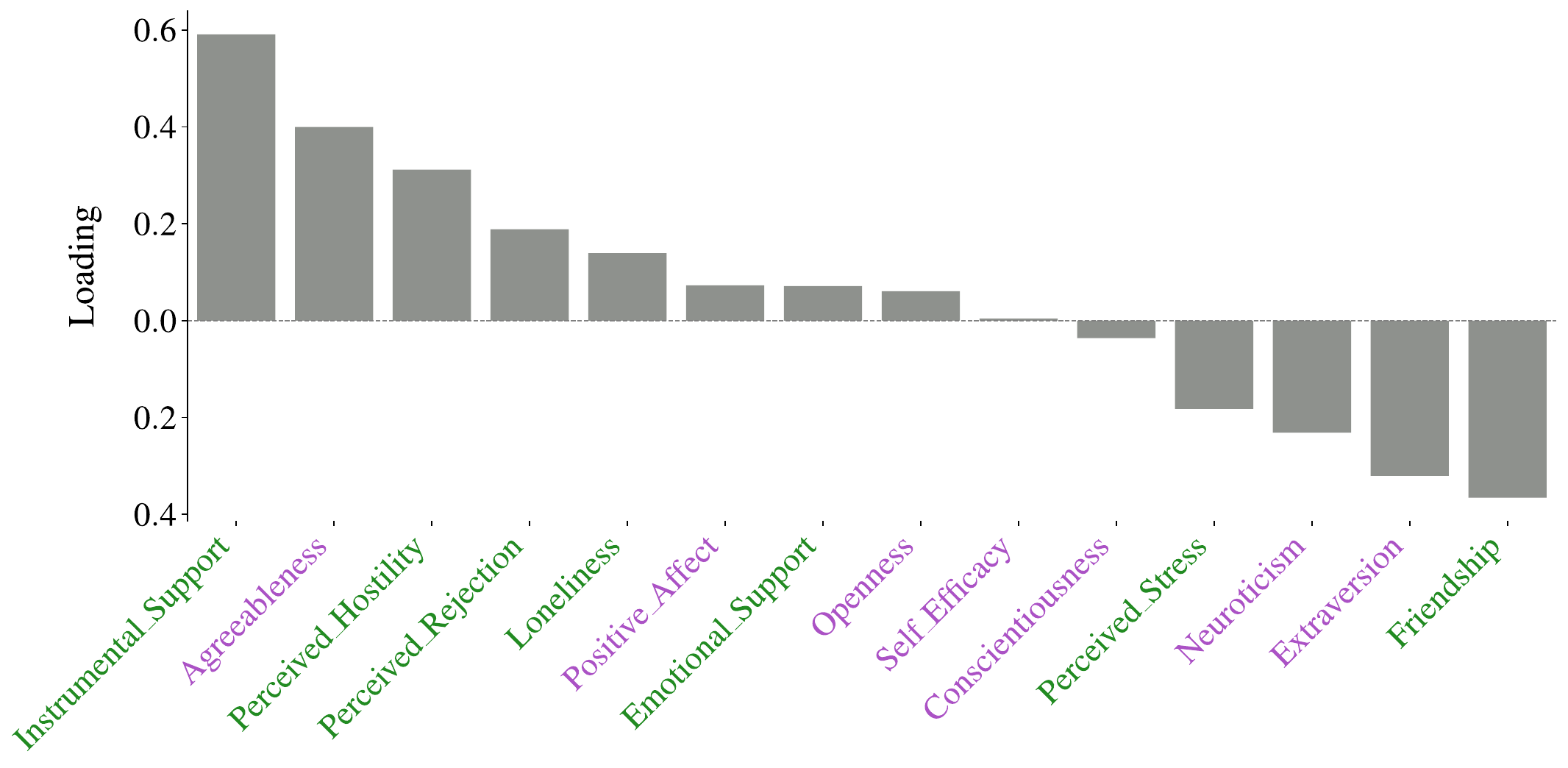}
        \caption{PC6}
        \label{fig:pc6}
    \end{subfigure}

    \caption{Loadings of the psychosocial measures on the first six principal components of the PCA applied to all psychosocial measures. Elements on the x-axis are ordered according to loading, with inner-self measures colored in purple and outer-self measures in green.}
    \label{fig:combined_pca_top_loadings}
\end{figure}

\begin{figure}[ht!]
    \centering

    \begin{subfigure}{0.32\linewidth}
        \centering
        \includegraphics[width=\linewidth]{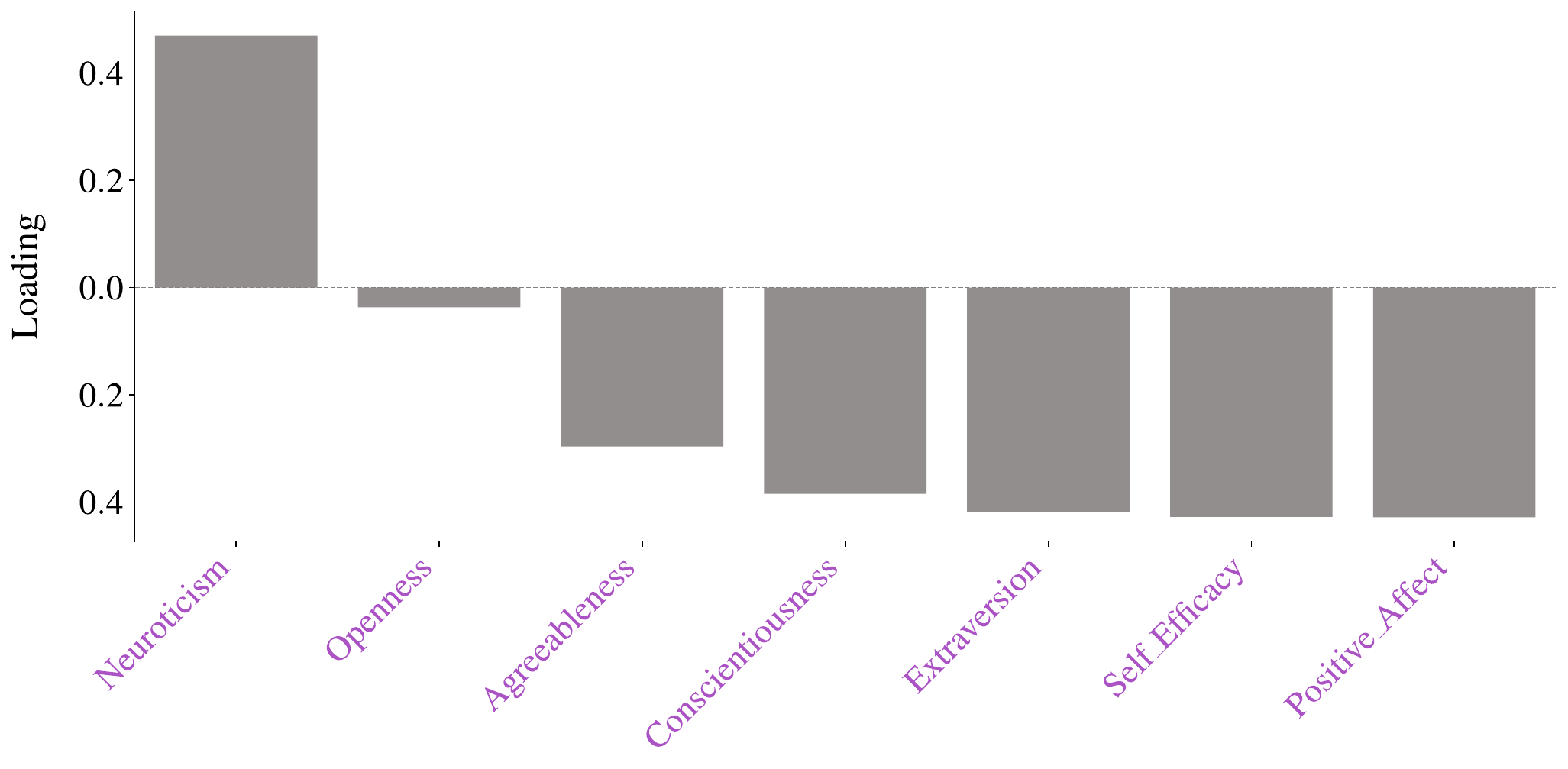}
        \caption{Inner PC1}
        \label{fig:inner_pc1}
    \end{subfigure}
    \begin{subfigure}{0.32\linewidth}
        \centering
        \includegraphics[width=\linewidth]{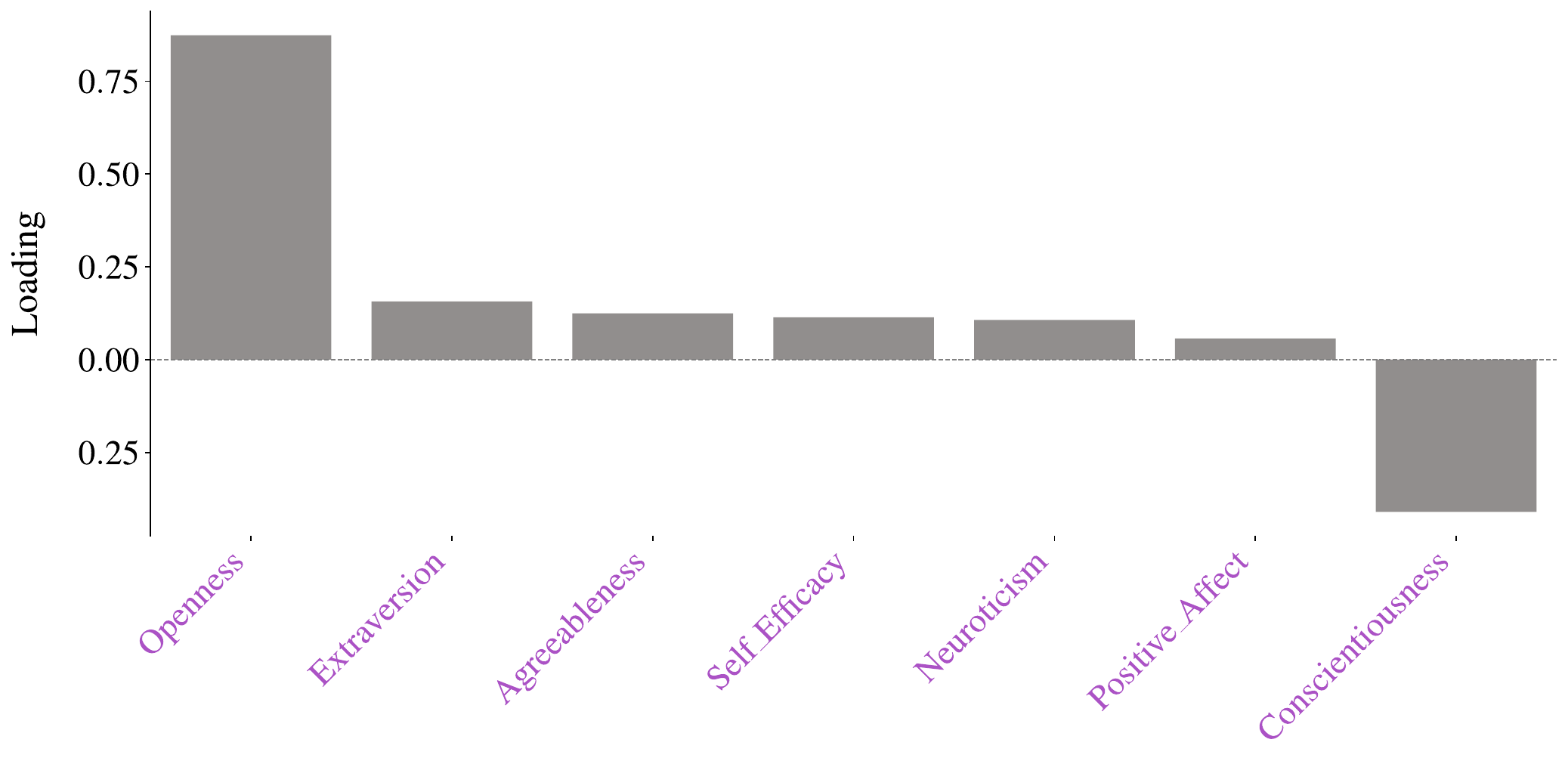}
        \caption{Inner PC2}
        \label{fig:inner_pc2}
    \end{subfigure}

    \vspace{0.4cm}

    \begin{subfigure}{0.32\linewidth}
        \centering
        \includegraphics[width=\linewidth]{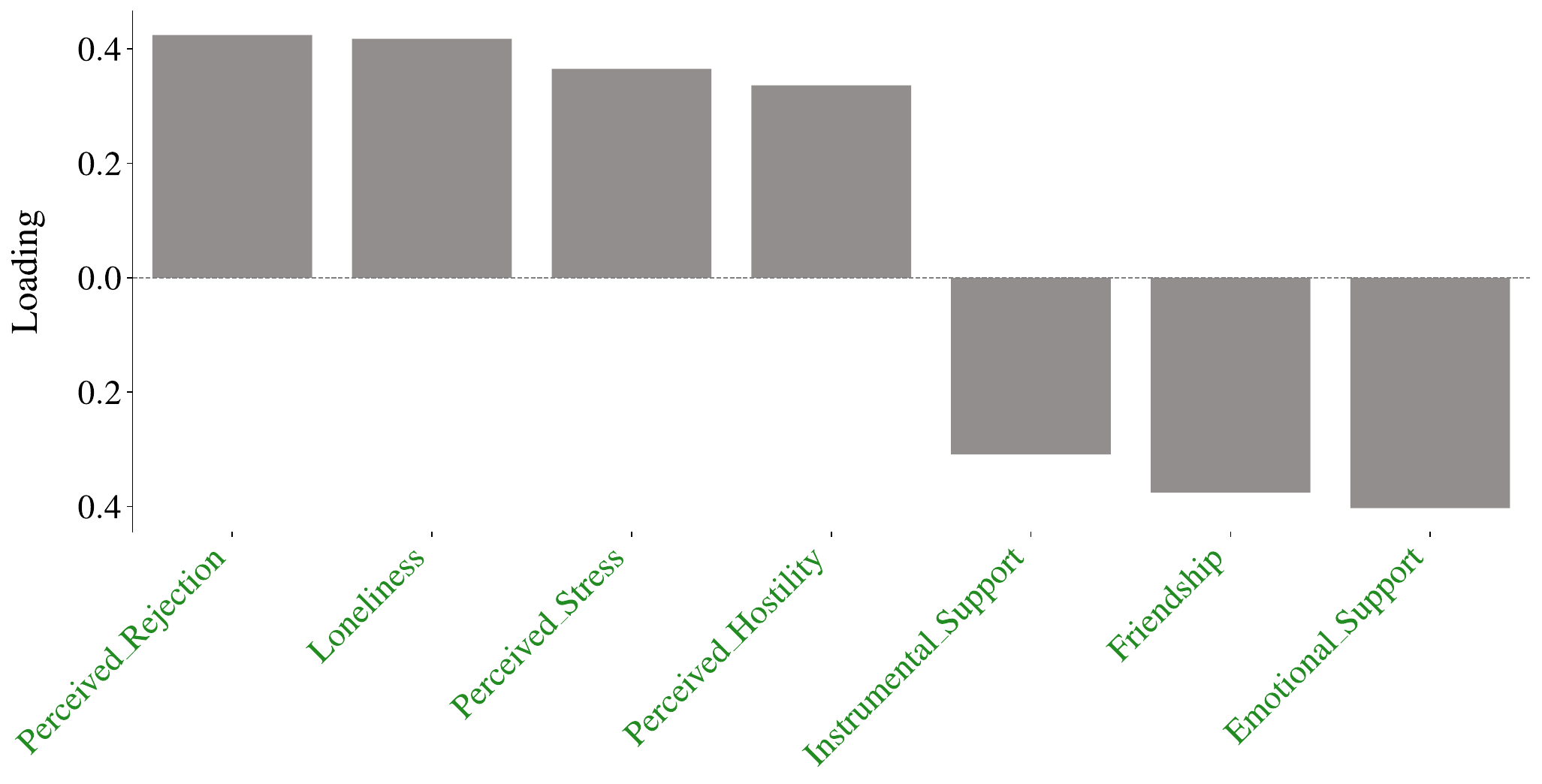}
        \caption{Outer PC1}
        \label{fig:outer_pc1}
    \end{subfigure}
    \begin{subfigure}{0.32\linewidth}
        \centering
        \includegraphics[width=\linewidth]{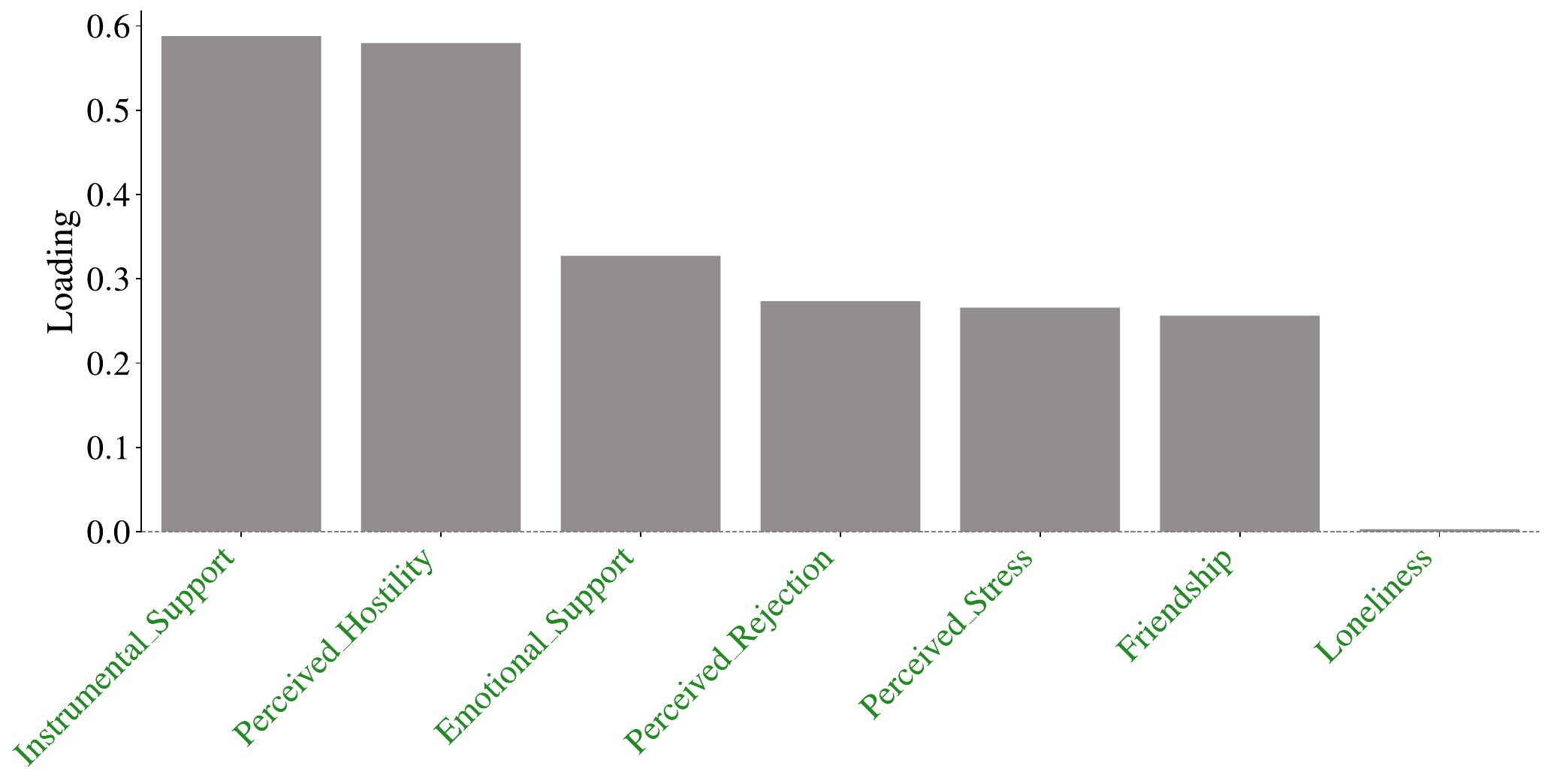}
        \caption{Outer PC2}
        \label{fig:outer_pc2}
    \end{subfigure}
  
    \caption{Loadings of the psychosocial measures on the first two principal components of the PCA applied to the inner-self measures (a-b) and outer-self measures (c-d). Elements on the x-axis are ordered according to loading, with inner-self measures colored in purple and outer-self measures in green.}
    \label{fig:pca_inner_outer_top_loadings}
\end{figure}

\newpage

\section{Optimal number of clusters}
\label{sec:appendixB}

To evaluate the optimal number of clusters, we compared simultaneously four validity indices: WCSS elbow criterion, Silhouette Score, Calinksi-Harabász index and Davies-Bouldin index. The most optimal number of clusters $k$, that is, the clustering which yields the most separated, cohesive and well-defined clusters, will be indicated by the higher values of WCSS elbow criterion, Silhouette Score, Calinksi-Harabász index; and the lower values of the Davies-Bouldin index. These criteria are most satisfied for $k=2$.

\begin{figure}[htbp]
    \centering 
    \includegraphics[width=0.8\textwidth]{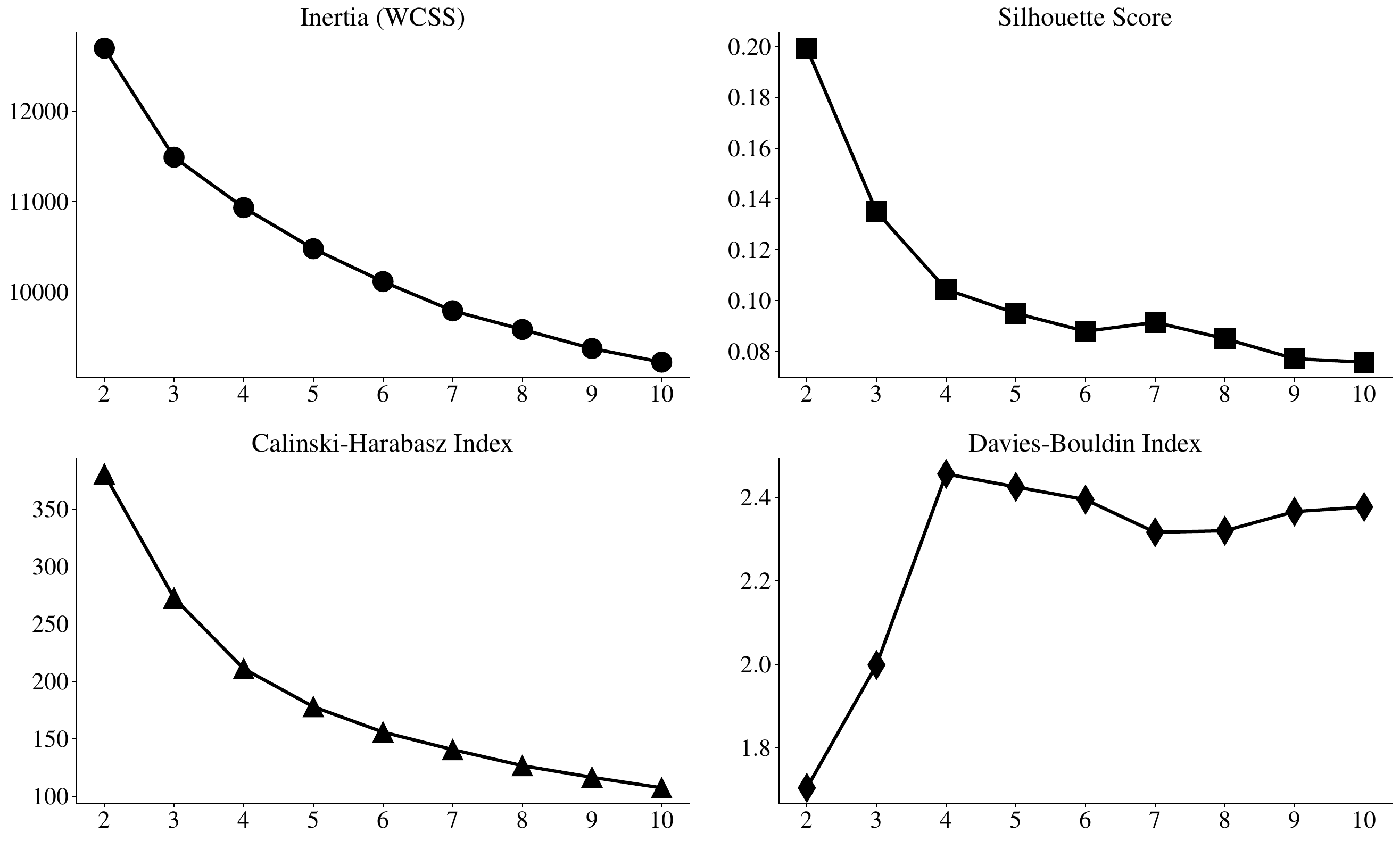}
    \caption{Results for the complementary validity indices for selection of the optimal $k$. WCSS elbow criterion (upper left), Silhouette Score (upper right), Calinski-Harabász index (lower left), Davies-Bouldin index (lower right).}
    \label{fig:cluster-metrics}
\end{figure}

\bibliographystyle{abbrvnat}
\newpage
\bibliography{references}  

\end{document}